\documentclass[draftcls, onecolumn, 10pt]{IEEEtran}
\usepackage{amssymb,amsmath}
\usepackage[dvips]{graphicx}
\usepackage{amsfonts}
\usepackage{subfigure}
\usepackage{booktabs,multirow}
\usepackage{color}
\usepackage{cite}
\usepackage[table]{xcolor}

\IEEEoverridecommandlockouts

\par
\begin{document}

\title{Wireless Fractal Cellular Networks}

\author{\normalsize
Xiaohu Ge$^1$,~\IEEEmembership{Senior~Member,~IEEE,} Yehong Qiu$^1$, Jiaqi Chen$^1$, Meidong Huang$^1$, \\Hui Xu$^2$, Jing Xu$^2$, Wuxiong Zhang$^2$,Yang Yang$^2$,~\IEEEmembership{Senior~Member,~IEEE,} \\Cheng-Xiang Wang$^3$,~\IEEEmembership{Senior~Member,~IEEE,} John Thompson$^4$,~\IEEEmembership{Fellow,~IEEE}\\
\vspace{0.70cm}
\small{
$^1$Department of Electronics and Information Engineering\\
Huazhong University of Science and Technology, Wuhan 430074, Hubei, P. R. China.\\
Email: \{xhge, yehong\_qiu, chenjq\_hust, u201113358\}@mail.hust.edu.cn\\
\vspace{0.2cm}
$^2$Key Lab of Wireless Sensor Network and Communication,  \\
Shanghai Research Center for Wireless Communications,  \\
Shanghai Institute of Microsystem and Information technology, \\
Chinese Academy of Sciences, Shanghai 200050, China\\
Email: \{hui.xu, jing.xu, wuxiong.zhang, yang.yang\}@wico.sh\\
\vspace{0.2cm}
$^3$Institute of Sensors, Signals and Systems, \\
School of Engineering \& Physical Sciences, \\
Heriot-Watt University, Edinburgh, EH14 4AS, UK.\\
Email: cheng-xiang.wang@hw.ac.uk\\
\vspace{0.2cm}
$^4$Institute for Digital Communications, \\
University of Edinburgh, Edinburgh, EH9 3JL, UK.\\
Email: john.thompson@ed.ac.uk\\
\vspace{0.2cm}
}
\thanks{\small{ Accepted by IEEE Wireless Communications.}}
}

\renewcommand{\baselinestretch}{1.2}
\thispagestyle{empty}
\maketitle
\thispagestyle{empty}
\setcounter{page}{1}\begin{abstract}
With the seamless coverage of wireless cellular networks in modern society, it is interesting to consider the shape of wireless cellular coverage. Is the shape a regular hexagon, an irregular polygon, or another complex geometrical shape? Based on fractal theory, the statistical characteristic of the wireless cellular coverage boundary is determined by the measured wireless cellular data collected from Shanghai, China. The measured results indicate that the wireless cellular coverage boundary presents an extremely irregular geometrical shape, which is also called a statistical fractal shape. Moreover, the statistical fractal characteristics of the wireless cellular coverage boundary have been validated by values of the Hurst parameter estimated in angular scales. The statistical fractal characteristics of the wireless cellular coverage boundary can be used to evaluate and design the handoff scheme of mobile user terminals in wireless cellular networks.

\end{abstract}

\IEEEpeerreviewmaketitle

\newpage
\section{Introduction}
It is estimated that 90\% of the world's population over 6 years old will have a mobile phone by 2020, i.e., most of the population will be covered by wireless cellular networks\cite{Ericson}. The coverage shape of a wireless cell is formed by the wireless cellular coverage boundary, which is connected by all of the farthest locations around a base station (BS). The farthest locations are also called wireless cellular coverage boundary points, where the received wireless signal power is equal to the minimum power threshold ${P_{min}}$ configured by the cellular network. An important challenge for wireless cellular network providers is to ensure that mobile users are seamlessly covered by adjacent BSs, especially those located at the edge of wireless cells\cite{Chen15}. Moreover, the user handoff between adjacent wireless cell signals depends on the wireless cellular coverage boundary in wireless cellular networks. Therefore, the shape of the wireless cellular coverage boundary is a critical metric for the design, deployment and optimization of wireless cellular networks.

Wireless cellular coverage shapes have been investigated for wireless cellular networks over the past few decades\cite{Andrews,Baccelli}.  Assuming that the propagation environment is free space and that the BS wireless signal is uniformly radiated in all directions, the wireless cellular coverage shape should be a circle with the BS located at the center in a two-dimensional plane\cite{Andrews}. When BSs are assumed to be uniformly deployed with equal distances in a wireless cellular network, the wireless cellular network service region can be split into multiple regular triangles, squares or regular hexagons that seamlessly cover the service region without overlaps. Considering that a regular hexagon is the closest to a circle among all candidate shapes, i.e., regular triangle, square and regular hexagon, a regular hexagon has been widely adopted as the wireless cellular coverage model in conventional wireless cellular networks\cite{Andrews}. With an increase in the density of BSs, existing studies have indicated that the performance of regular hexagon wireless cellular networks deviates from the performance of real wireless cellular networks\cite{Baccelli}. Based on measured data, the locations of BSs can be approximated by a Poisson Point Process distribution for wireless cellular networks\cite{Baccelli}. Moreover, the wireless cell boundaries, which are obtained through the Delaunay Triangulation method by connecting the perpendicular bisector lines between each pair of BSs, split the wireless cellular network service region into multiple irregular polygons that correspond to different wireless cellular coverage areas. This stochastic and irregular topology creates the need for a so-called Poisson-Voronoi tessellation (PVT) method \cite{Andrews}. However, the impact of wireless signal propagation environments on the wireless cell boundary is not considered in the PVT random wireless cellular network models. Moreover, to simplify system models, the path loss fading of a wireless signal in the PVT network model is assumed to be equal in all directions if the distances between receivers and the BS are equal. This assumption ignores the anisotropy of path loss fading in real wireless signal propagation environments. Moreover, conventional geometric segmentation methods used to form wireless cell boundaries, such as the PVT method, result in a smooth wireless cell boundary at small scales. However, the measured cellular data indicates that the PVT method cannot provide an accurate estimation of real wireless cellular coverage shapes \cite{Goldsmith93}.

The wireless cellular coverage boundary is not smooth at small scales because the wireless signal fading in real environments is affected by the electromagnetic radiation, the atmospheric environment, the weather status, the obstacle distribution, and diffraction and scattering effects in different propagation directions. Considering the irregular distribution of buildings in urban environments, electromagnetic waves are absorbed, reflected, scattered and diffracted in different directions. Therefore, in urban environments, wireless signals transmitted by BSs undergo different amounts of attenuation and fading in different directions before arriving at the users. The existing study in \cite{Haenggi08} validated that the probability density function (pdf) of the interference exhibits a heavy-tailed characteristic. Moreover, the traffic load of cellular networks has been demonstrated to manifest the self-similar characteristic which also conduces to the heavy-tailed distribution of traffic load \cite{Lilith05}. Essentially, several effects such as the wireless signal attenuation, the network traffic and the interference caused by adjacent BSs may affect the shape of wireless cellular coverage boundary. As a consequence, wireless cellular coverage boundaries will present extremely irregular shapes at small scales for real wireless cellular networks. However, it is difficult to describe extremely irregular wireless cell boundaries using conventional Euclidean geometry methods.

As an important extension of the conventional Euclidean geometry theory, fractal geometry theory describes geometric shapes between extreme geometric orders and full chaos \cite{Mandelbrot}. Based on the measured wireless cellular data, we utilize the typical wireless signal propagation model and the least squares method to estimate the path loss coefficient and shadow fading. In this case, real wireless signal propagation environments are focused and other potential roots, such as the network traffic and the interference resulting in heavy-tailed characteristic of wireless cellular coverage boundary are ignored in this study. We report that the real wireless cellular coverage boundary is a non-smooth boundary in urban environments. Furthermore, utilizing fractal geometry theory, the real wireless cellular coverage boundary has statistical fractal characteristics at angular scales and real wireless signal propagation environments conduce to statistical fractal characteristics of the wireless cellular coverage boundary in angular scales. The statistical fractal is not an exact fractal that can be denoted by an exact fractal expression \cite{Mandelbrot69}. Compared with an exact fractal, the statistical fractal is more suitable for describing geometric shapes in the real world. The quantization of a statistical fractal is typically estimated by the value of the Hurst parameter\cite{Hurst}.  Three typical statistical estimators, i.e., the periodogram method, the rescaled adjusted range statistic (R/S) method and the variance-time analysis method, are utilized to estimate the value of the Hurst parameter for real wireless cellular coverage boundaries. The estimated results indicate that the real wireless cellular coverage boundary has the statistical fractal characteristic at angular scales. Oppositely, a comparison of the results demonstrates that a mathematically derived wireless cellular coverage boundary does not have statistical fractal characteristics at angular scales. Although the experimental measurement in this paper is carried out in cellular networks, the analysis results reflect the coverage characteristic of wireless communications considering wireless signal propagation environments. Therefore, our results can also be used for other wireless communication scenarios, such as WLANs.

\section{Measured and derived wireless cellular coverage shapes}
The wireless signal power received at a mobile user terminal is measured by a continuous wave test signal method that is widely used to evaluate wireless propagation environments\cite{Erceg}. The measurement solution in this paper is configured as follows: a BS equipped with an omnidirectional antenna is located at Pingjiang Road, Shanghai, China. The BS transmits wireless signals at a fixed frequency of 2.6 GHz and a fixed transmission power of 38 dBm. The detailed BS configuration parameters are illustrated in Table \uppercase\expandafter{\romannumeral1}. The received wireless signal power is measured by a mobile user terminal equipped with an omnidirectional antenna that moves along a specified route, as shown in Fig. 1. The measurement data was collected on May 15, 2014. The specified route passes through office buildings, residential houses and green belts in Shanghai. The mobile user terminal moved around the cellular coverage region to measure the received wireless signal power and the corresponding global positioning system (GPS) data, which facilitated the estimation of the distance between the mobile user terminal and BS.

\begin{figure*}[!h]
\begin{center}
\includegraphics[width=5in]{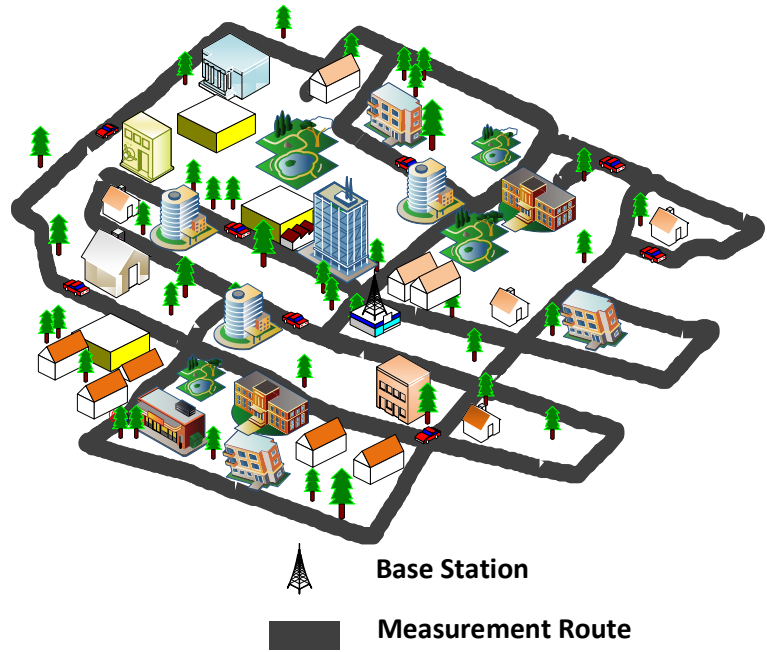}
\caption{BS location and measurement route. The BS location is denoted by a BS icon, and the gray line is the measurement route.}\label{Fig1}
\end{center}
\end{figure*}

\begin{table*}[!h]
\setlength{\abovecaptionskip}{0mm}
\centering
\caption{Base Station configuration}
\label{tab:1}
{\small
\begin{tabular}{cccccc}\toprule
\noalign{\smallskip}
\textbf{Measurement } & \textbf{GPS location} & \textbf{Transmiss-} & \textbf{Feeder loss}&\textbf{Antenna trans-} & \textbf{Antenna rece- }\\
\textbf{environment}&\textbf{of BS}&\textbf{ion power}&\textbf{at the antenna}&\textbf{mission gain (Tx)}&\textbf{ive gain (Rx)}\\ \midrule
\noalign{\smallskip}\noalign{\smallskip}
{Pingjiang Road,  } &{Latitude=31.202252 } & {38 dBm}&{0.5 dB}&{12 dBi}&{3 dBi} \\
{Shanghai, China} & {Longitude=121.451055} &  \\ \bottomrule
\noalign{\smallskip}
\end{tabular}
}
\end{table*}

In wireless communications, the wireless signal fading is typically classified in two parts: large-scale fading, which includes path loss fading and shadow fading, and small-scale fading, which includes multipath fading. In practical wireless signal measurement applications, the received wireless signal power is averaged over several wavelengths to eliminate the multipath fading effect\cite{Lee}. In our wireless signal measurements, the wireless signal power is averaged over 40 wavelengths by a mobile user terminal. In this case, the small-scale fading is ignored due to the averaged multipath fading effect \cite{Erceg}. Moreover, the path loss fading is denoted as ${d^{-\gamma}}$ , where ${\gamma}$ is the path loss coefficient and $d$ is the distance between a receiver and a BS. The shadow fading is denoted as $\psi$ . For the derived wireless cellular coverage regions, the path loss fading is assumed to be equal in all directions of the derived wireless cellular coverage regions when the distance between a receiver and a BS is the same.  Based on the measured wireless cellular data, the average path loss coefficient ${\gamma}$ is estimated using the least squares method. In addition, the shadow fading is assumed to follow a log-normal distribution\cite{Ge}. Without loss of generality, the shadow fading is assumed to follow a log-normal distribution with a mean of $\mu=0$ dB and a standard deviation of $\sigma=4$ dB for the mathematically derived wireless cellular coverage regions \cite{Jakes}. However, for real wireless cellular coverage regions, the path loss fading and shadow fading is not the same in different propagation directions and depends on real propagation paths.

 In Fig. 2a, the transmission wireless signal power presents a power peak at the BS location and then uniformly attenuates in all propagation directions with increasing distance. Moreover, the transmission wireless signal power varies smoothly with increasing distance, especially at locations far away from the BS. When the minimum received wireless signal power threshold is configured as ${P_r=-110}$ dBm, the corresponding equal power curve at -110 dBm of the received wireless signal is plotted to form the derived wireless cellular coverage region, as shown in Fig. 2b. The derived wireless cellular coverage boundary appears as an amoeba around the BS.

\begin{figure*}[!h]
\setlength{\abovecaptionskip}{-1mm}
\centering
\begin{tabular}{cc}
\begin{minipage}[t]{3in}
\includegraphics[width=3in]{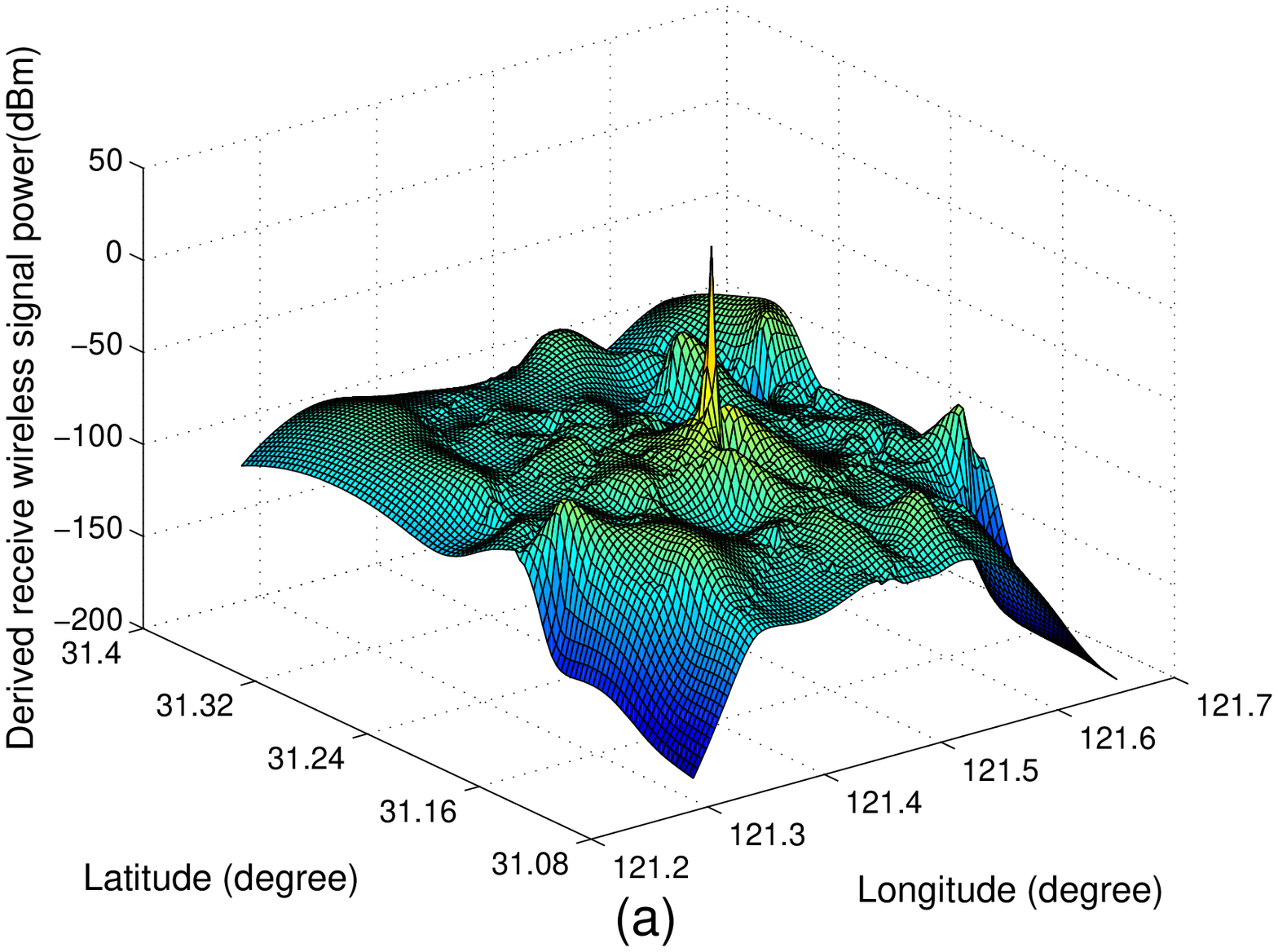}
\end{minipage}
\begin{minipage}[t]{3in}
\includegraphics[width=3in]{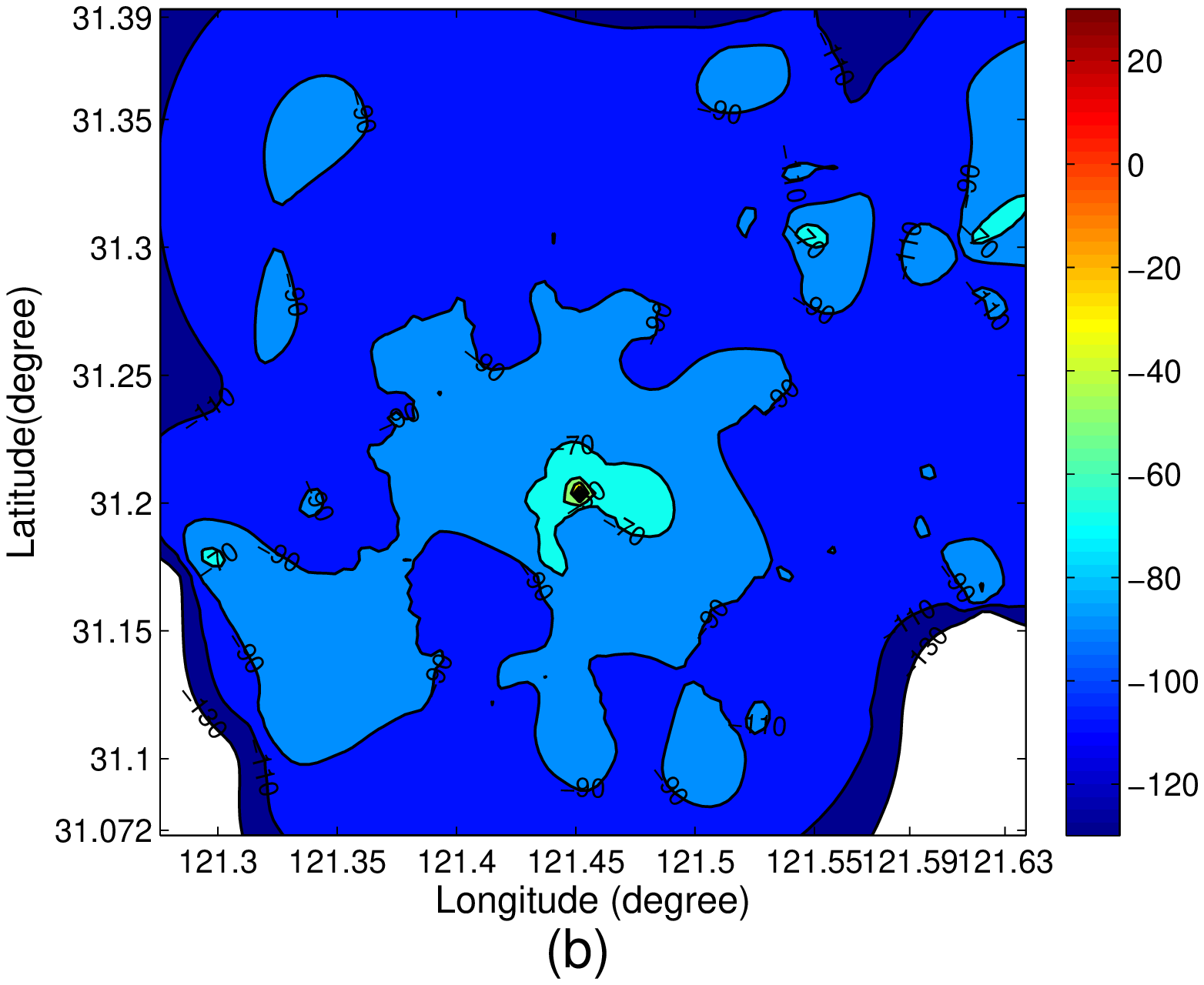}
\end{minipage}\\
\begin{minipage}[t]{3in}
\includegraphics[width=3in]{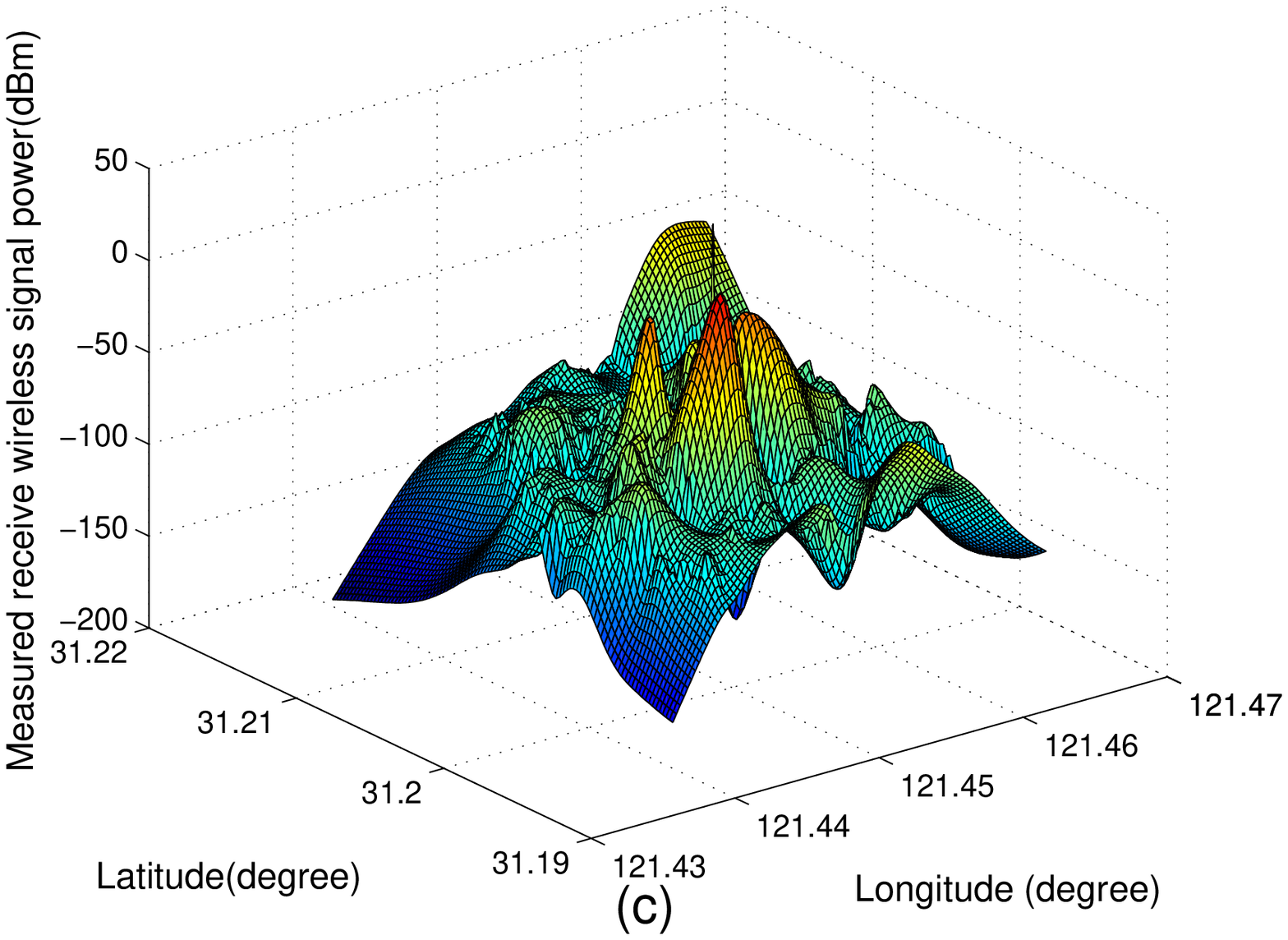}
\end{minipage}
\begin{minipage}[t]{3in}
\includegraphics[width=3in]{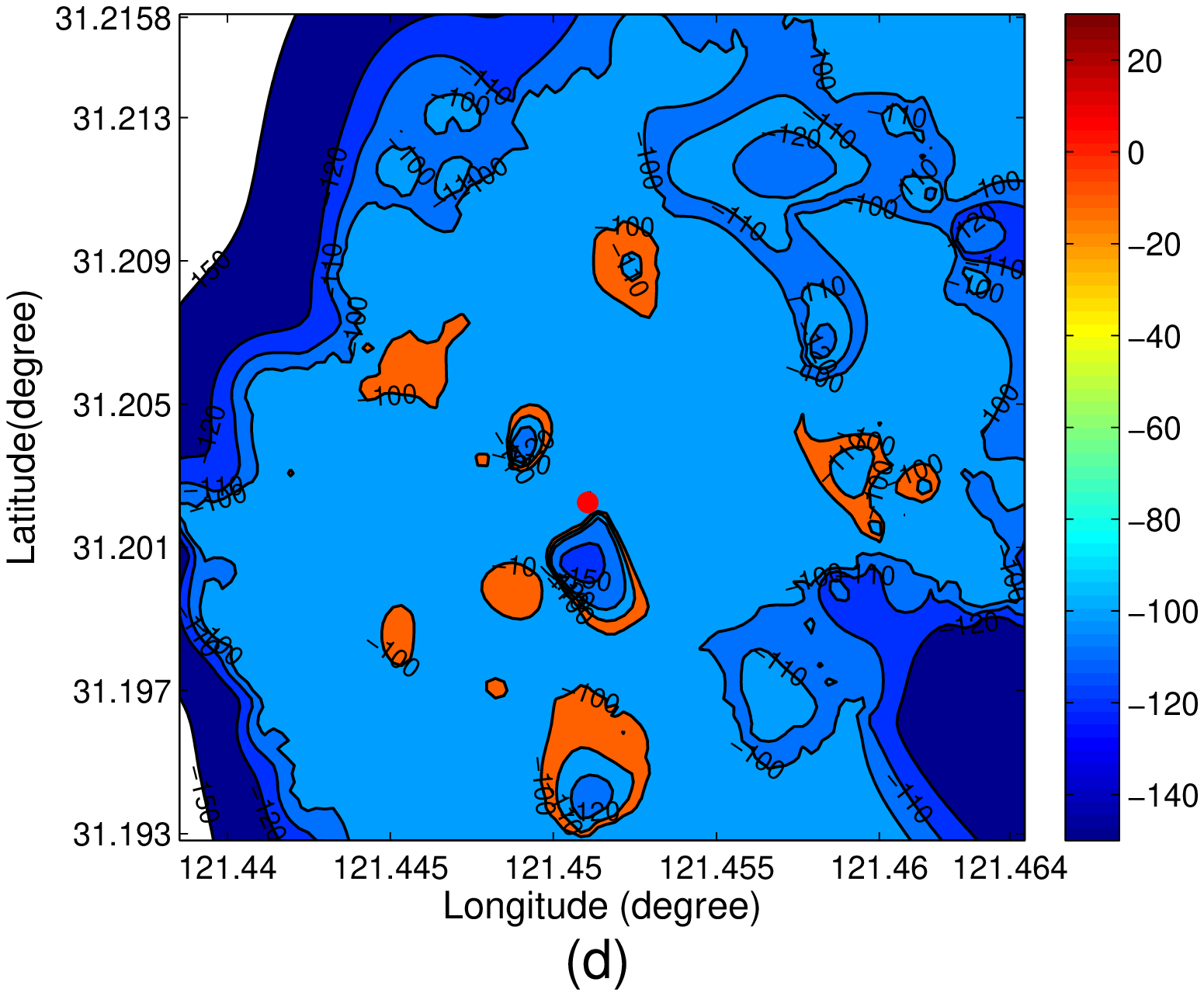}
\end{minipage}
\end{tabular}
\caption{Derived and measured received wireless signal power figures. The bright red regions denote the highest receive wireless signal power, and the ultramarine regions denote the lowest receive wireless signal power. The received wireless signal power decreases with the color changing from bright red to ultramarine. (a) Received wireless signal power level based on the derived wireless cellular coverage. (b) Received wireless signal equal power curve based on the derived wireless cellular coverage. The black line is the received equal power liner level; the same color region indicates that the value of the wireless signal power is between two power values denoted by two black lines. (c) Received wireless signal power level based on the measured wireless cellular coverage. (d) Received wireless signal equal power curve based on the measured wireless cellular coverage. The black line is the received equal power liner level, and the same color region indicates that the value of wireless signal power is between two power values denoted by two black lines. }\label{Fig2}
\end{figure*}

 Considering the non-uniform distribution of different sizes of office buildings and large obstacles in real propagation paths, the path loss fading should present anisotropy in different propagation directions in urban environments. In other word, the path loss fading values at different locations are different even though these locations have the same distance to the BS in real wireless propagation environments. Furthermore, the path loss coefficient is not a constant in different propagation directions and depends on real propagation paths. In Figs. 2c and 2d, the path loss coefficients of different directions are estimated using the measured wireless cellular data in corresponding directions. Moreover, the shadow fading is estimated using a least squares method for measured wireless cellular coverage regions. Fig. 2c shows the transmission wireless signal power measured from real propagation environments. In Fig. 2c, the transmission wireless signal power presents a power peak at the BS location and then non-uniformly attenuates in different propagation directions with increasing distance. Clear phenomena can be observed, with a "mountain top" and a "mountain valley" around the power peak, as shown in Fig. 2c. This result implies that the path loss fading exhibits the expected anisotropy in different propagation directions. When the minimum received wireless signal power threshold is configured as ${P_r=-110}$ dBm, the corresponding received wireless signal equal power curve with -110 dBm is plotted to form the measured wireless cellular coverage region, as shown in Fig. 2d. The measured wireless cellular coverage boundary does not have the appearance of an amoeba around the BS and presents extreme irregularity at small scales. Therefore, it is difficult to describe the real wireless cellular boundary using conventional Euclidean geometry methods.

\section{Measured and derived wireless cellular coverage boundaries}
To quantitatively analyze the irregularity of wireless cellular coverage boundaries, the received wireless signal equal power line is discretized for evaluating the geometry characteristics in small scales. To obtain discrete wireless cellular coverage boundary points, the measured wireless cellular coverage region is partitioned into 120 sections centered at the BS and ${3^\circ}$ angular width. Moreover, the path loss coefficient is assumed to be identical in a section propagation environment when the section angle is sufficiently small, such as when the section angle is less than or equal to ${3^\circ}$.
Based on the testing route circling around the BS, testing points are obtained and distributed in the same section with different distances from the BS. Compared with a mix of no line of sight (NLOS) and line of sight (LOS) path loss model, a general path loss model would be an average of the path loss coefficient in the corresponding section. Moreover, the general path loss model does not change the relationship with other sections in a cell. Hence, a general path loss model is adopted in this study by
${P_{{r_d}}} = {P_{{r_{{d_0}}}}} - 10\gamma {\log _{10}}(\frac{d}{{{d_0}}}) - \psi $\cite{Andrews}, where  ${P_{{r_d}}}$ is the measured received wireless signal power at a mobile user terminal whose distance from the BS is $d$ , ${P_{{r_{{d_0}}}}}$ is the measured received wireless signal power at a reference location whose distance from the BS is ${d_0}$, $\gamma $ and $\psi $ are the path loss coefficient and shadow fading, respectively. Based on the wireless signal propagation model and measured received wireless signal power of testing points in the $k - th$  section, the path loss coefficient ${\gamma _k}$ and shadow fading ${\psi _k}$ are estimated by a least squares method for the $k - th$  section of the measured wireless cellular coverage region. For details, the values of $({P_{{r_{{d_0}}}}} - {P_{{r_d}}})$ versus $10{\log _{10}}(\frac{d}{{{d_0}}})$ are first plotted in a Euclidean coordinate system. The wireless signal propagation model is fitted by a least squares line which is formed by measured data collected from the $k - th$ section. As a result, the asymptotic slope of the least squares fits the path loss coefficient ${\gamma _k}$ and the asymptotic intercept is the shadow fading ${\psi _k}$ in the $k - th$ section of the measured wireless cellular coverage region. When the received wireless signal power threshold at the wireless cellular coverage boundary is configured as ${P_{{r_\phi }}} = {P_{\min }} =  - 110$ dBm and the reference location ${d_0}$  is configured as the $i - th$ testing point in the $k - th$ section, the distance ${\phi _{k,i}}$ between the discrete boundary point and the BS is calculated by the wireless signal propagation model based on the estimated path loss coefficient ${\gamma _k}$  and shadow fading ${\psi _k}$  in the $k - th$  section of the measured wireless cellular coverage region. When all sections of the wireless cellular coverage region are measured, the distance series $\phi $ of the measured wireless cellular coverage boundary is obtained for further statistical analysis.
The measured wireless cellular coverage boundary is plotted in Fig. 3a when all discrete boundary points are connected. The average path loss coefficient is calculated by the total measured wireless cellular data for the derived wireless cellular coverage region. Moreover, the average path loss coefficient is used for every section in the derived wireless cellular coverage region, and the shadow fading is estimated by a log-normal distribution. Furthermore, a discrete boundary point is derived by the wireless channel propagation model for a section in the derived wireless cellular coverage region. The derived wireless cellular coverage boundary is plotted in Fig. 3b when all discrete boundary points are connected. Comparing Figs. 3a and 3b, the measured wireless cellular coverage boundary presents a large-scale fluctuation, i.e., a bursty characteristic, while the derived wireless cellular coverage boundary is a smooth circle.

\begin{figure*}[!h]
\setlength{\abovecaptionskip}{0mm}
\center
\begin{tabular}{cc}
\begin{minipage}[t]{2in}
\includegraphics[width=2in]{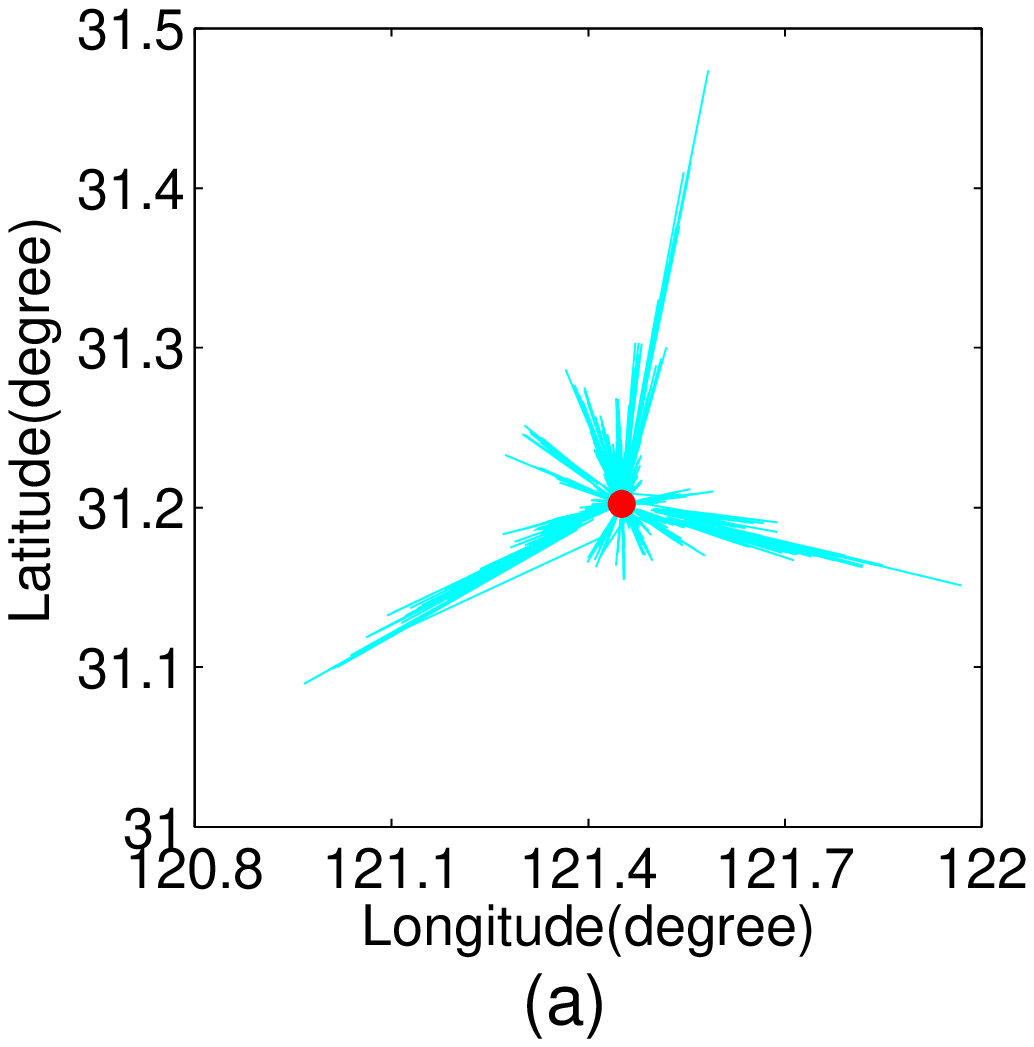}
\end{minipage}
\begin{minipage}[t]{2in}
\includegraphics[width=2in]{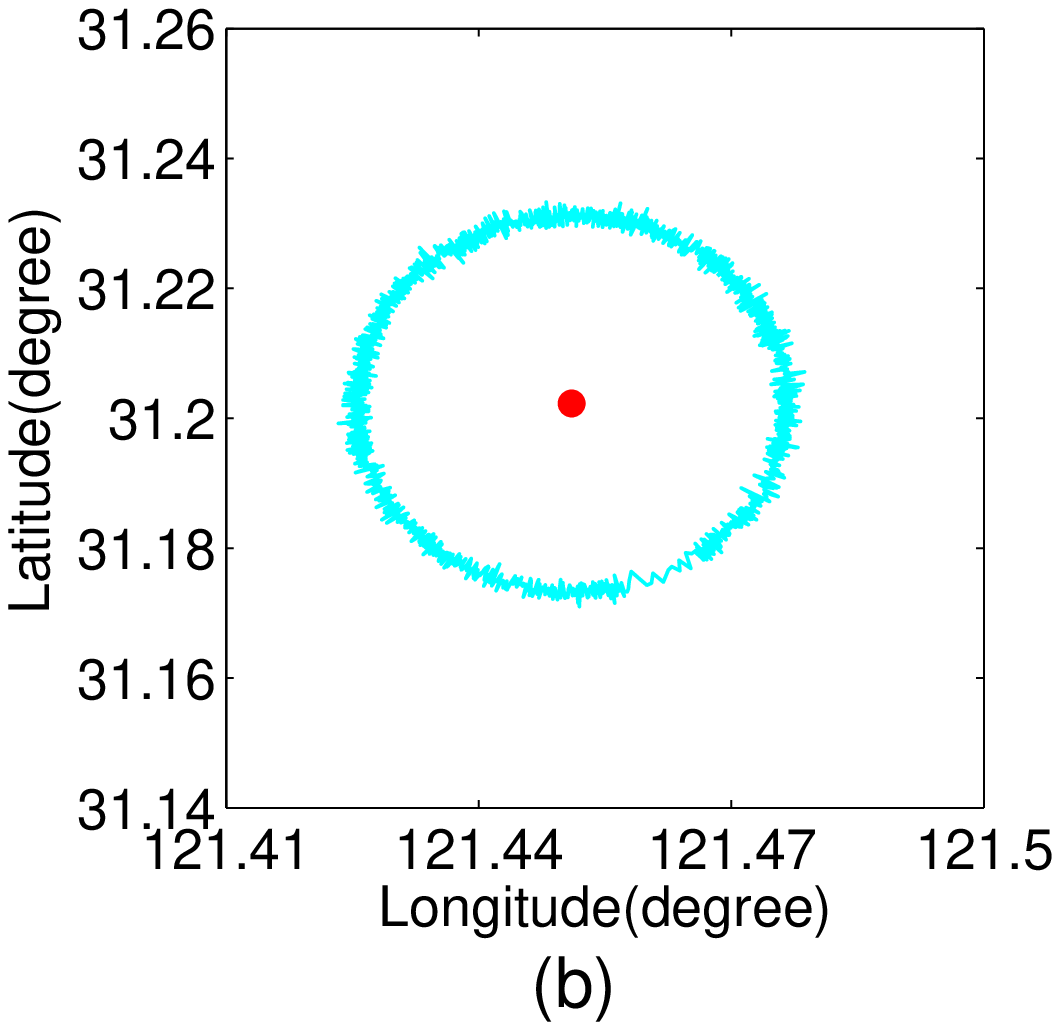}
\end{minipage}\\
\begin{minipage}[t]{2in}
\includegraphics[width=2in]{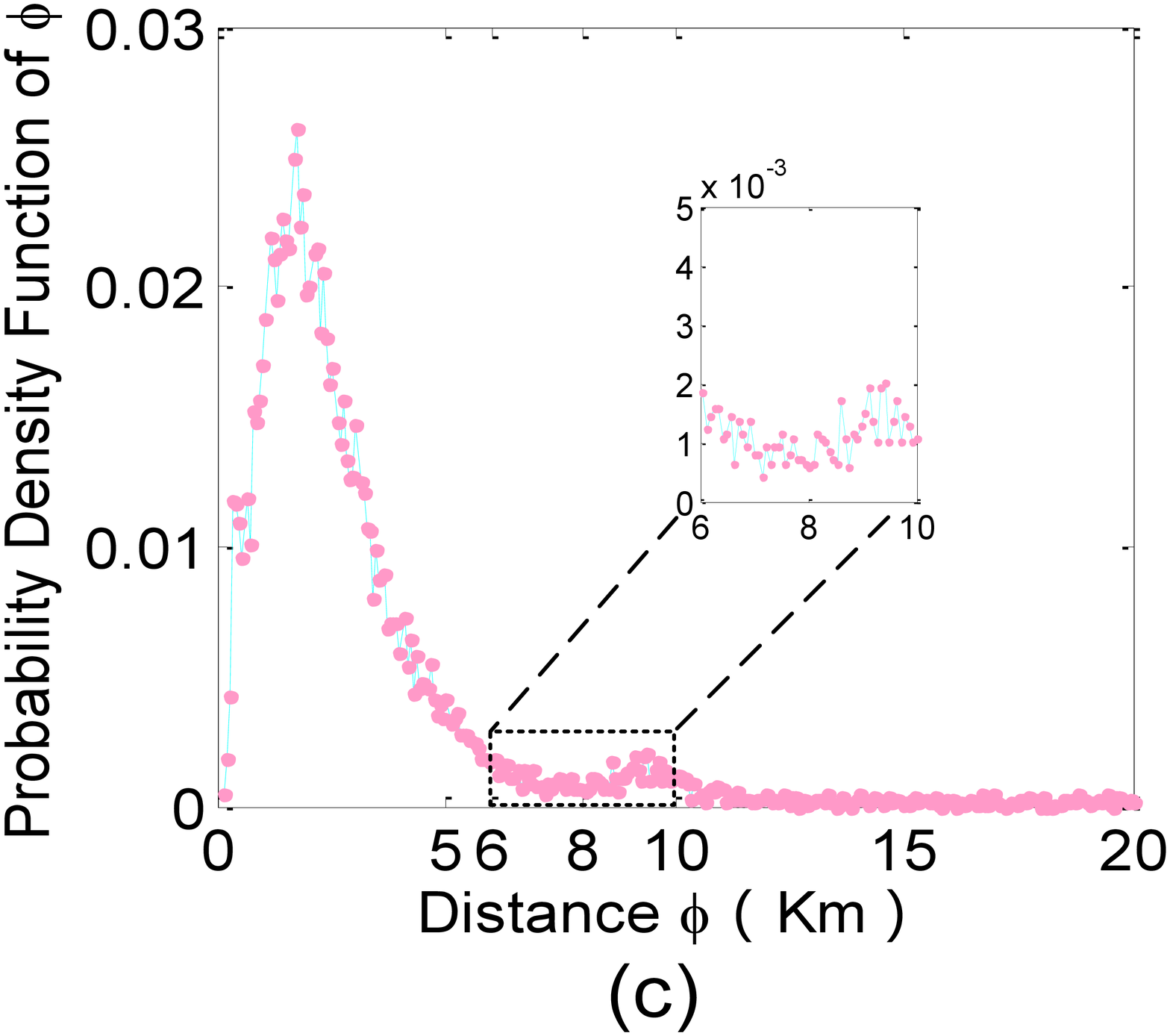}
\end{minipage}
\begin{minipage}[t]{2in}
\includegraphics[width=2in]{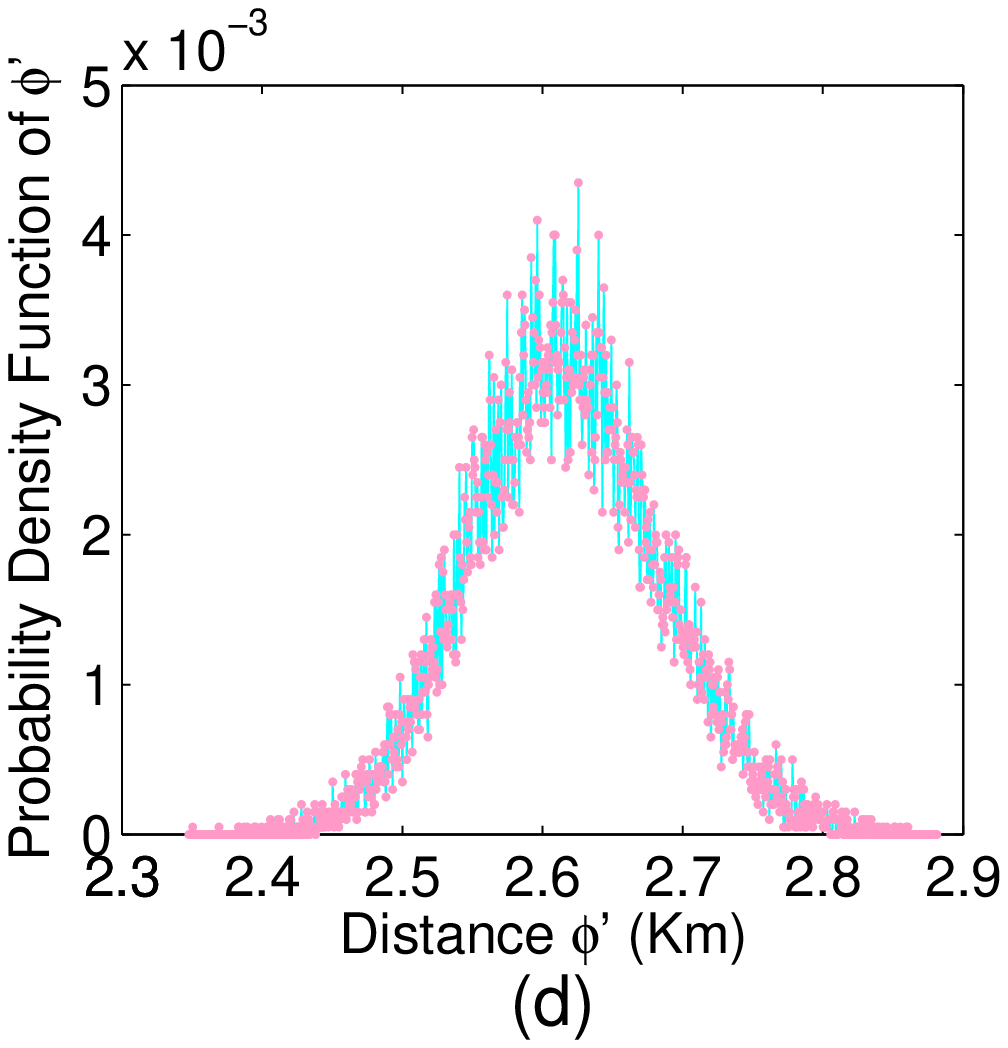}
\end{minipage}\\
\begin{minipage}[t]{1.8in}
\includegraphics[width=1.8in]{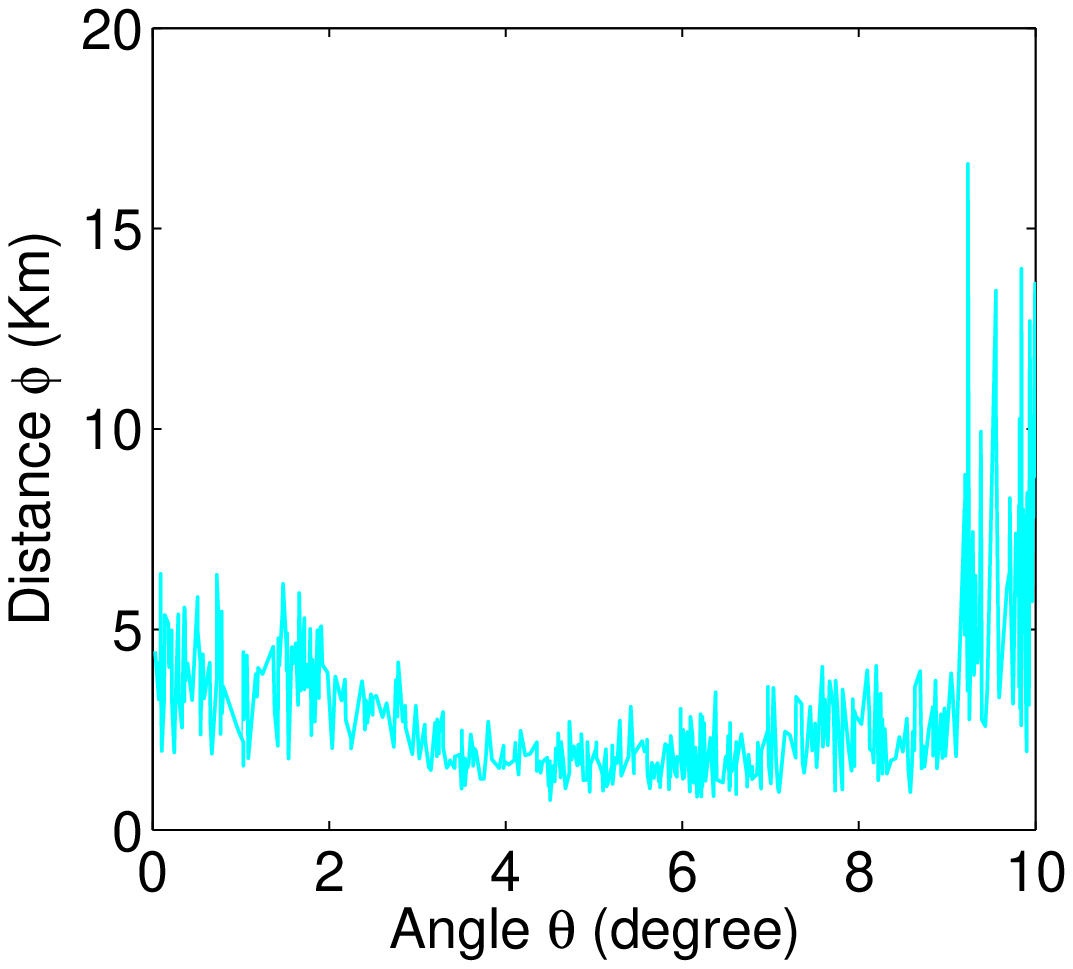}
\end{minipage}
\begin{minipage}[t]{1.8in}
\includegraphics[width=1.8in]{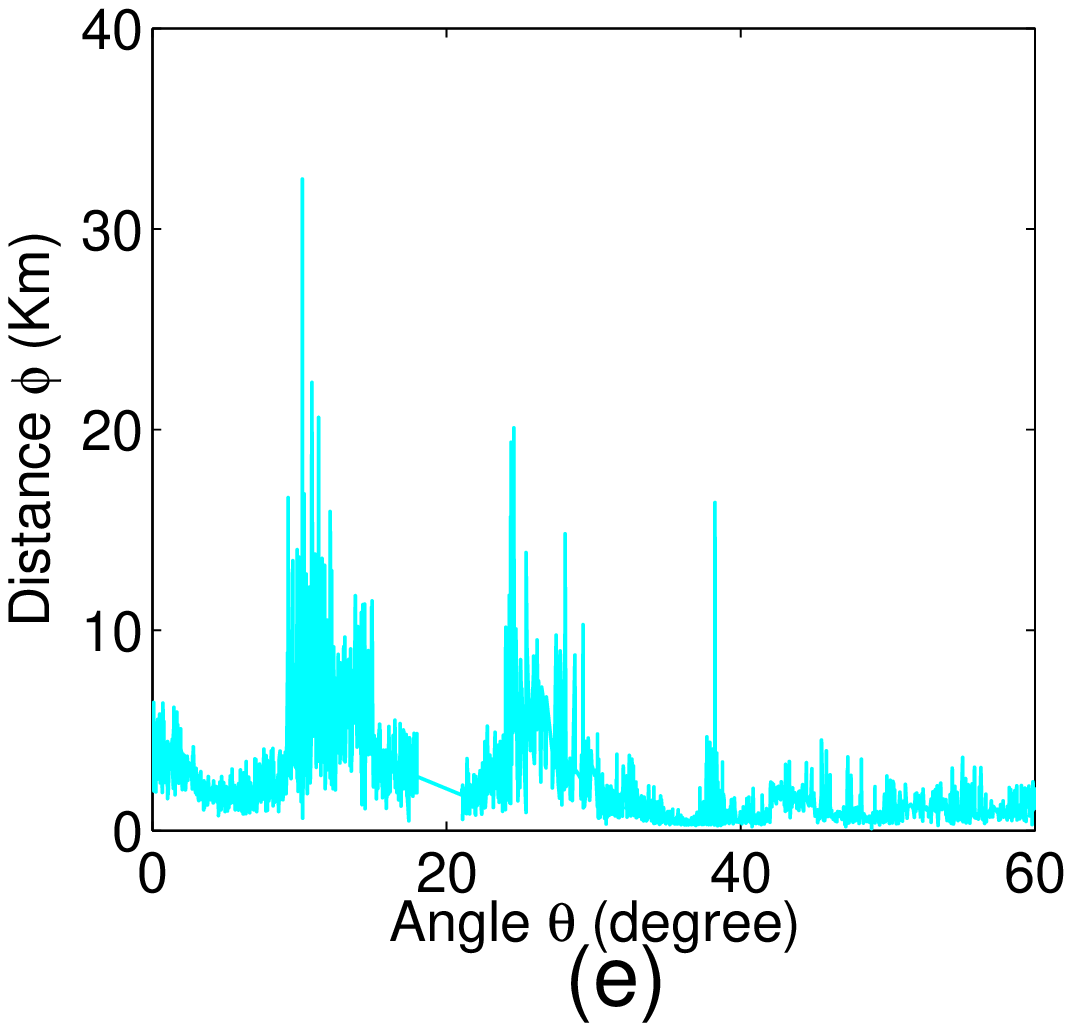}
\end{minipage}
\begin{minipage}[t]{1.8in}
\includegraphics[width=1.8in]{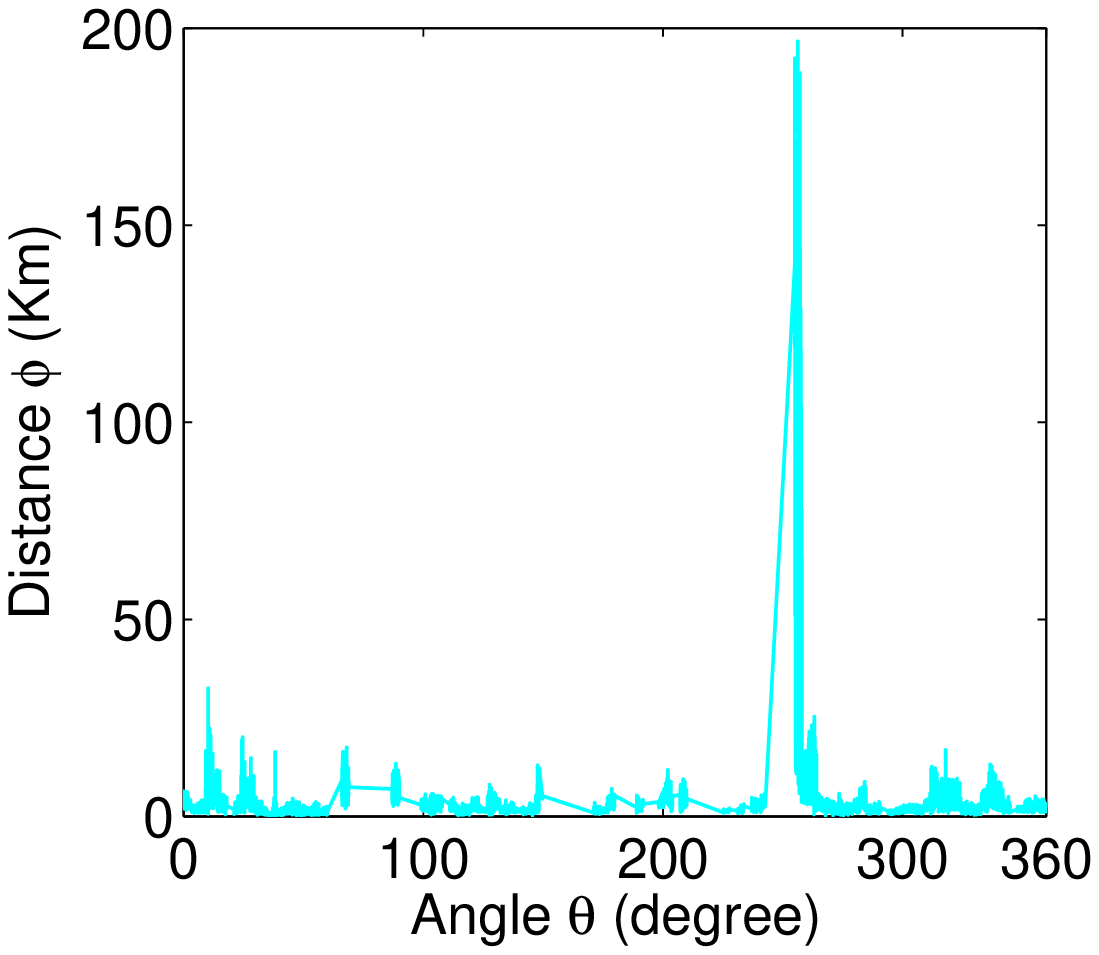}
\end{minipage}\\
\begin{minipage}[t]{1.8in}
\includegraphics[width=1.8in]{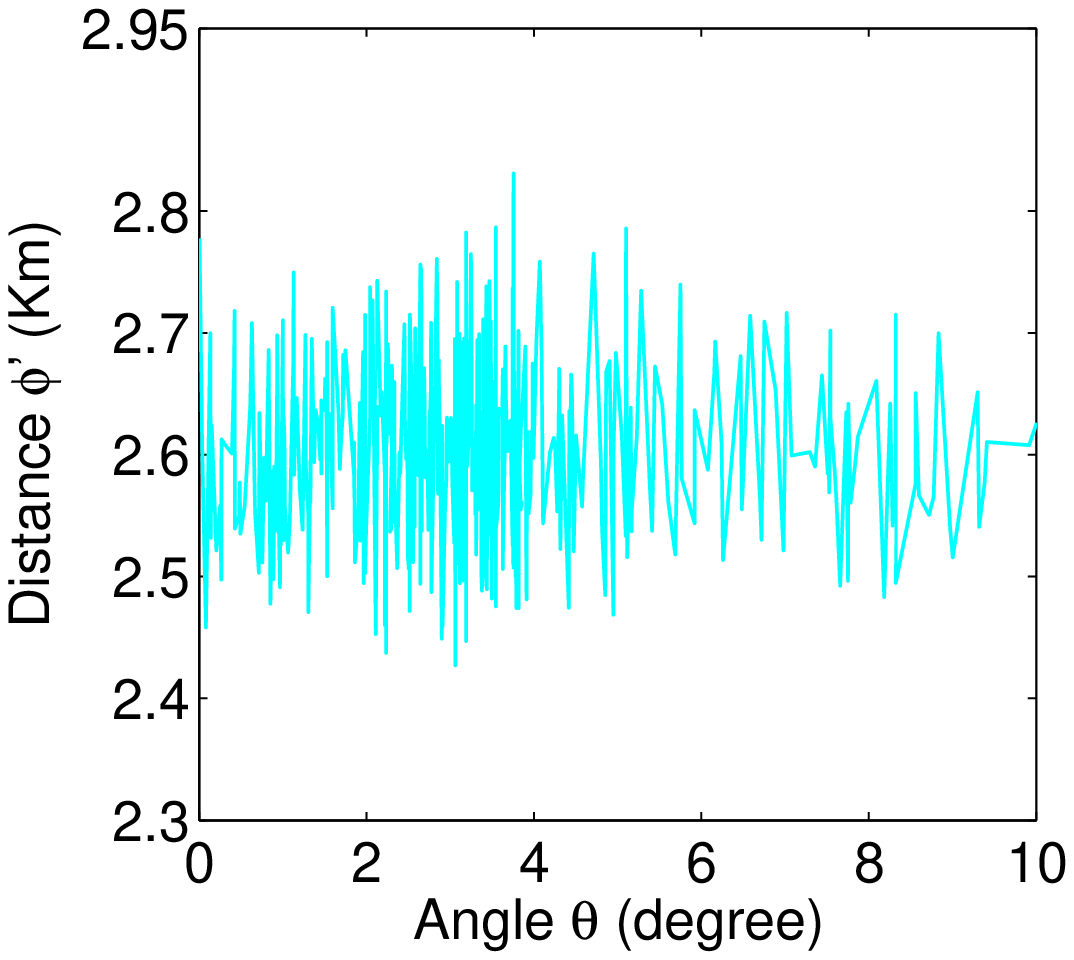}
\end{minipage}
\begin{minipage}[t]{1.8in}
\includegraphics[width=1.8in]{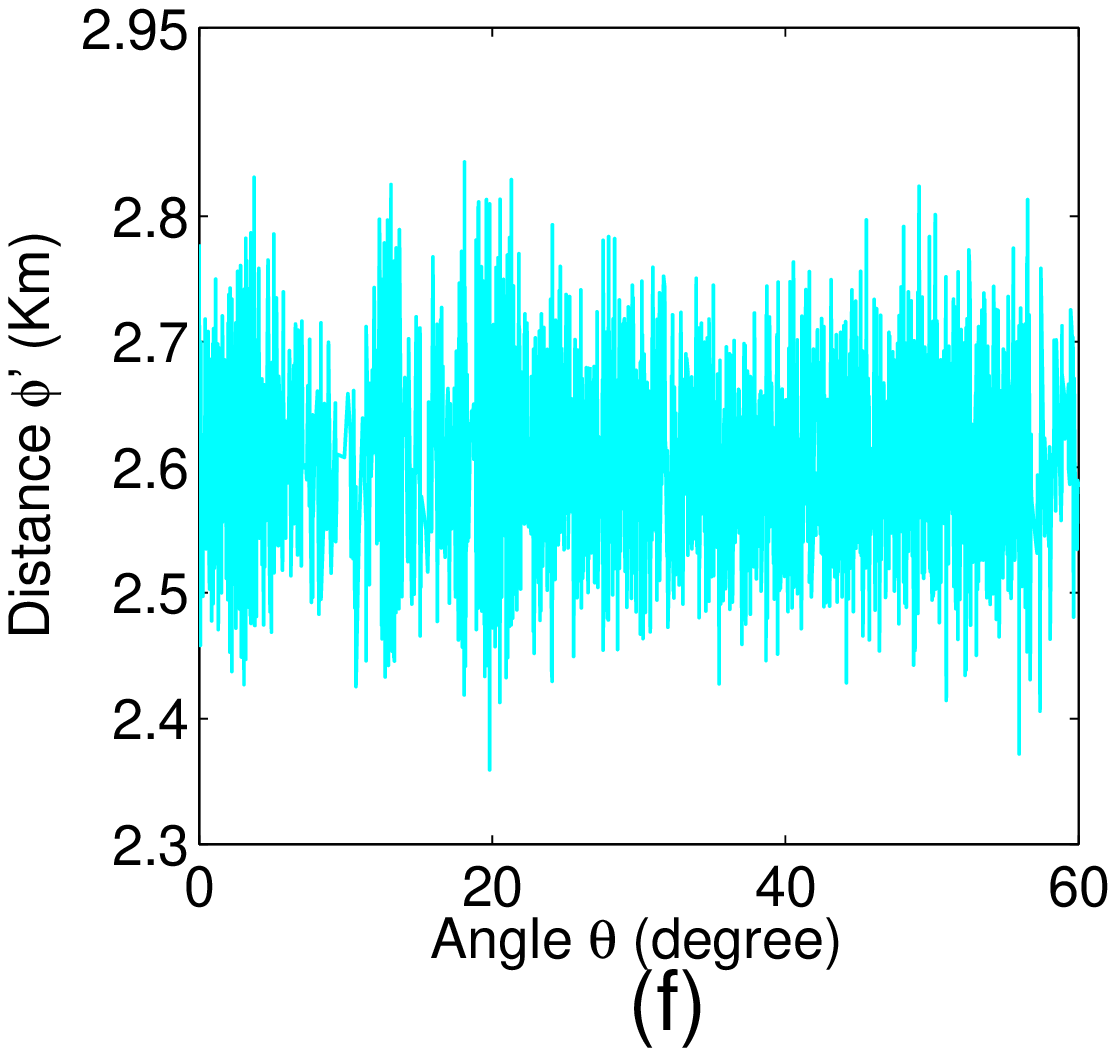}
\end{minipage}
\begin{minipage}[t]{1.8in}
\includegraphics[width=1.8in]{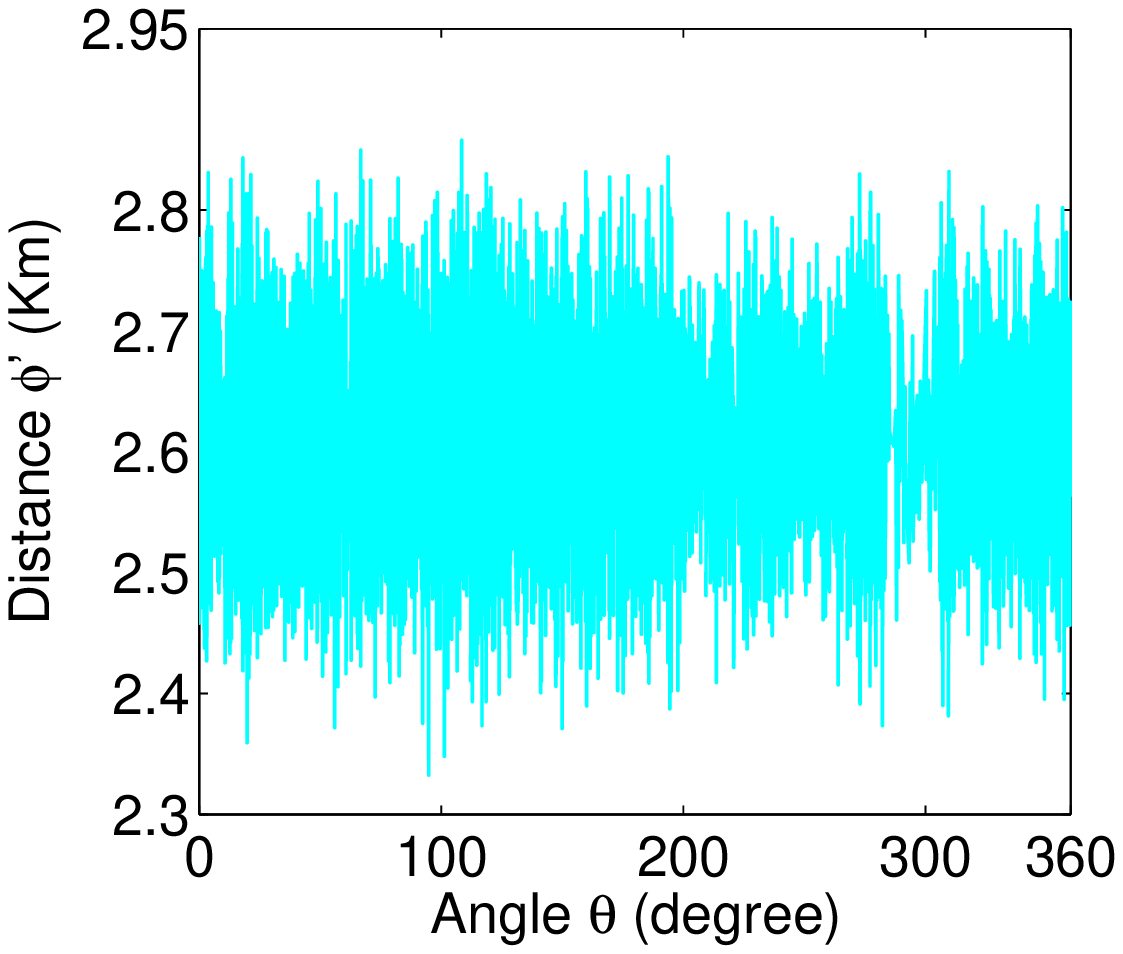}
\end{minipage}
\end{tabular}
\caption{ Measured and derived wireless cellular coverage boundaries. (a) Measured wireless cellular coverage shape; the red point is the BS, and the blue curve is the measured wireless cellular coverage boundary. (b) Derived wireless cellular coverage shape; the red point is the BS, and the blue curve is the derived wireless cellular coverage boundary. (c) PDF of the distance $\phi $ , where the pink points are the statistical probability points generated from the measured wireless cellular coverage boundary and the blue curve is the contact line in discrete statistical probability points. (d) PDF of the distances $\phi'$ , where the pink points are the statistical probability points generated from the derived wireless cellular coverage boundary and the blue curve is the contact line in discrete statistical probability points. (e) Distances $\phi$ between the discrete boundary point and BS with respect to the angle $\theta$ in the measured wireless cellular coverage boundary. (f) Distances $\phi'$ between the discrete boundary point and BS with respect to the angle $\theta$ in the derived wireless cellular coverage boundary.}\label{Fig3}
\end{figure*}

To analyze the statistical characteristic of the wireless cellular coverage boundary, let ${\phi}$ and ${{\phi}'}$ be the distances between the BS and a discrete boundary point in the measured wireless cellular coverage boundary and derived wireless cellular coverage boundary, respectively. Figs. 3c and 3d are the probability density function (PDF) of the distances ${\phi}$ and ${{\phi}'}$, respectively. Based on results in Fig. 3d, the shape of the PDF of the distance ${{\phi}'}$ is a typical shape of a Gaussian distribution. Compared with the shape in Fig. 3d, the shape of the PDF of the distance ${\phi}$ presents a heavy-tailed characteristic in Fig. 3c. The heavy-tailed characteristic of the PDF of the distance ${\phi}$ implies that some small probability events, such as some discrete boundary points that are far away from the BS, cannot be ignored in forming the distribution of measured wireless cellular coverage boundary. Hence, the distribution of measured wireless cellular coverage boundary is a non-Gaussian distribution.

Considering the heavy-tailed and bursty characteristics in the analysis of the measured wireless cellular coverage boundary, we investigate the measured wireless cellular coverage boundary using fractal theory. Unlike conventional fractal studies at temporal scales and spatial scales \cite{Leland}, the fractal study of the measured wireless cellular coverage boundary is analyzed in angular scales in this paper. Without loss of generality, an angle denoted as ${\theta}$ is between the east direction line and a given line that is crossed with a discrete boundary point and the BS. Fig. 3e illustrates distances ${\phi}$ between discrete boundary points and the BS with respect to the angle ${\theta}$ for the measured wireless cellular coverage boundary. The peak range of the distance for the measured wireless cellular coverage boundary exhibits burstinesses when the angle is restricted by ${0^\circ}{\sim}{10^\circ}$.  When the angle ${\theta}$ is extended from ${0^\circ}{\sim}{10^\circ}$ to ${0^\circ}{\sim}{60^\circ}$, i.e., the angle scale is zoomed in six times, the distances ${\phi}$  clearly exhibits burstiness, and the peak value of the distance is 34 km. When the angle ${\theta}$ is extended from ${0^\circ}{\sim}{60^\circ}$ to ${0^\circ}{\sim}{360^\circ}$, i.e., the angle scale is zoomed in six times again, the distances ${\phi}$ still exhibits clear burstiness, and the peak value of the distance is 200 km. Fig. 3e shows that the burstiness of distances at different angular scales of the measured wireless cellular coverage boundary cannot be smoothed by zooming in to angular scales, i.e., there always exists burstiness of the distances at all angular scales of the measured wireless cellular coverage boundary. This phenomenon is called a fractal or self-similarity phenomenon in angular scales \cite{Mordechai}. Fig. 3f illustrates distances ${{\phi}'}$  between the discrete boundary points and BS with respect to the angle ${\theta}$ for the derived wireless cellular coverage boundary. The peak range of the distance exhibits burstiness for the derived wireless cellular coverage boundary when the angle is restricted to the range of ${0^\circ}{\sim}{10^\circ}$. When the angle ${\theta}$ is extended from ${0^\circ}{\sim}{10^\circ}$ to ${0^\circ}{\sim}{60^\circ}$, i.e., the angular scale is zoomed in six times, then the peak range of the distance ${{\phi}'}$ is nearly smooth. When the angle ${\theta}$ is extended from ${0^\circ}{\sim}{60^\circ}$ to ${0^\circ}{\sim}{360^\circ}$, i.e., the angular scale is zoomed in six times again, the peak range of the distance  ${{\phi}'}$ is fully smooth. Fig. 3f indicates that the burstiness of distances at small angular scales can be smoothed with increases in the angular scales in the derived wireless cellular coverage boundary. Hence, there is no fractal or self-similarity phenomenon in the angular scales for the mathematically derived wireless cellular coverage boundary.

\section{Fractal evaluation of the measured wireless cellular coverage boundary}
No exact fractal phenomenon exists in the real world. Most of the fractal phenomena observed in the real world only have the statistical fractal characteristic [6-7].  Statistical fractal random processes present the spectral density power-law behavior and the slowly decaying variance characteristic in the frequency and time domains, respectively. Moreover, the statistical fractal characteristic of random processes is evaluated by the Hurst parameter, which can be estimated using three typical methods\cite{Ge04}:
\begin{enumerate}
\item	
 	The periodogram method. This method plots the logarithm of the spectral density of a series versus the logarithm of frequencies. The Hurst parameter can be estimated by ${H=\frac{1}{2}(1+\alpha)}$  where $\alpha$ is the slope in the log-log plot. The series has a statistical fractal character if ${0.5<H<1}$ .
\item
 	The rescaled adjusted range statistic (R/S) method. For a random process ${X_i}$, the partial sum is denoted by ${Y{{\left(n\right)}}=\sum\limits_{i=1}^{n}{X_i}}$, and sample variance is denoted by
 ${S^2}\left( n \right) = \frac{{\sum\nolimits_{i = 1}^n {\left( {{X_i} - \frac{Y\left( n \right)}{n}}\right)}^2}}{n},\ {\rm{ }}n \ge 1$. Furthermore, the R/S statistic is defined as $\frac{{R\left( n \right)}}{{S\left( n \right)}} = \frac{{\mathop {\max }\limits_{0 \le t \le n} \left(0, {Y\left( t \right) - \frac{t}{n}Y\left( n \right)} \right) - \mathop {\min }\limits_{0 \le t \le n} \left(0, {Y\left( t \right) - \frac{t}{n}Y\left( n \right)} \right)}}{{s(n)}},\ {\rm{ }}n \ge 1$. A log-log plot of the R/S statistic versus the number of points of the aggregated series should be a straight line with the slope being an estimation of the Hurst parameter. The random process ${X_i}$ is statistical fractal if the value of the Hurst parameter $H$ is in the interval $\left( {0.5,1.0} \right)$.
\item
 	 The variance-time analysis method involves the definition of an aggregated series ${X^{(m)}}$, using different block sizes ${m}$. The log-log plot of the sample variance versus the aggregation level should be a straight line with the slope $\beta $ in the interval $\left( {0,1} \right)$ if the data are statistical fractal. In this case, $H = 1 - \frac{\beta }{2}$.
\end{enumerate}

Let ${X} = \{ {\phi_i},\ i = 1,2, \cdots N\}$ and ${X'} = \{ {{\phi}'}_i,\ i = 1,2, \cdots M\}$ be two independent random processes, where $N$ and $M$ is the number of discrete measured and derived wireless cellular coverage boundary points, ${\phi_i}$ and ${{\phi_i}'}$ are distances between the $i - th$ discrete boundary point and BS in the measured wireless cellular coverage boundary and derived wireless cellular coverage boundary, respectively. Fig. 4a illustrates the values of the Hurst parameter estimated from the measured wireless cellular coverage boundary data using the periodogram method, R/S method and variance-time analysis method. In the left figure of Fig. 4a, the periodogram method is used to estimate the value of the Hurst parameter based on the spectral density $I(w)$ of the random process ${X} = \{ {\phi_i},\ i = 1,2, \cdots N\}$ in a log-log plot. When the frequency value $w$ approaches zero, the spectral density presents a low-power decaying behavior in the log-log plot. Utilizing a least squares method, the slope, i.e., the decay rate of the spectral density in the log-log plot, is estimated as $\alpha {\rm{ = }}0.8026$. Furthermore, the value of the Hurst parameter is calculated by $H = \frac{1}{2}(1 + \alpha ) = 0.9013$. In the middle figure of Fig. 4a, the R/S method is used to estimate the value of the Hurst parameter by the R/S statistic of the random process ${X} = \{ {\phi_i},\ i = 1,2, \cdots N\}$ in a log-log plot. The R/S statistic increases linearly with increases in the length of series $n$ in a log-log plot. Utilizing a least squares method, the Hurst parameter, i.e., the slope in the log-log plot, is estimated as $H = 0.8898$ . In the right figure of Fig. 4a, the variance-time analysis method is used to estimate the value of the Hurst parameter by the variance of the aggregated series of the random process ${X} = \{ {\phi_i},\ i = 1,2, \cdots N\}$ in a log-log plot. The variance of the aggregated series presents a slow decaying characteristic with increases in the aggregation level $m$ in a log-log plot. Utilizing a least squares method, the slope of the sample variance is estimated as $\beta {\rm{ = }}0.2000$ in the log-log plot. Furthermore, the value of the Hurst parameter is calculated to be $H = 1 - \frac{\beta }{2} = 0.9000$. Based on the results from the three Hurst parameter estimators, the value of the Hurst parameter estimated from the measured wireless cellular coverage boundary is clearly larger than 0.5 and is approximately $H \approx 0.9$.

\begin{figure*}[!h]
\setlength{\abovecaptionskip}{0mm}
\centering
\begin{tabular}{cc}
\begin{minipage}[t]{1.8in}
\includegraphics[width=1.8in]{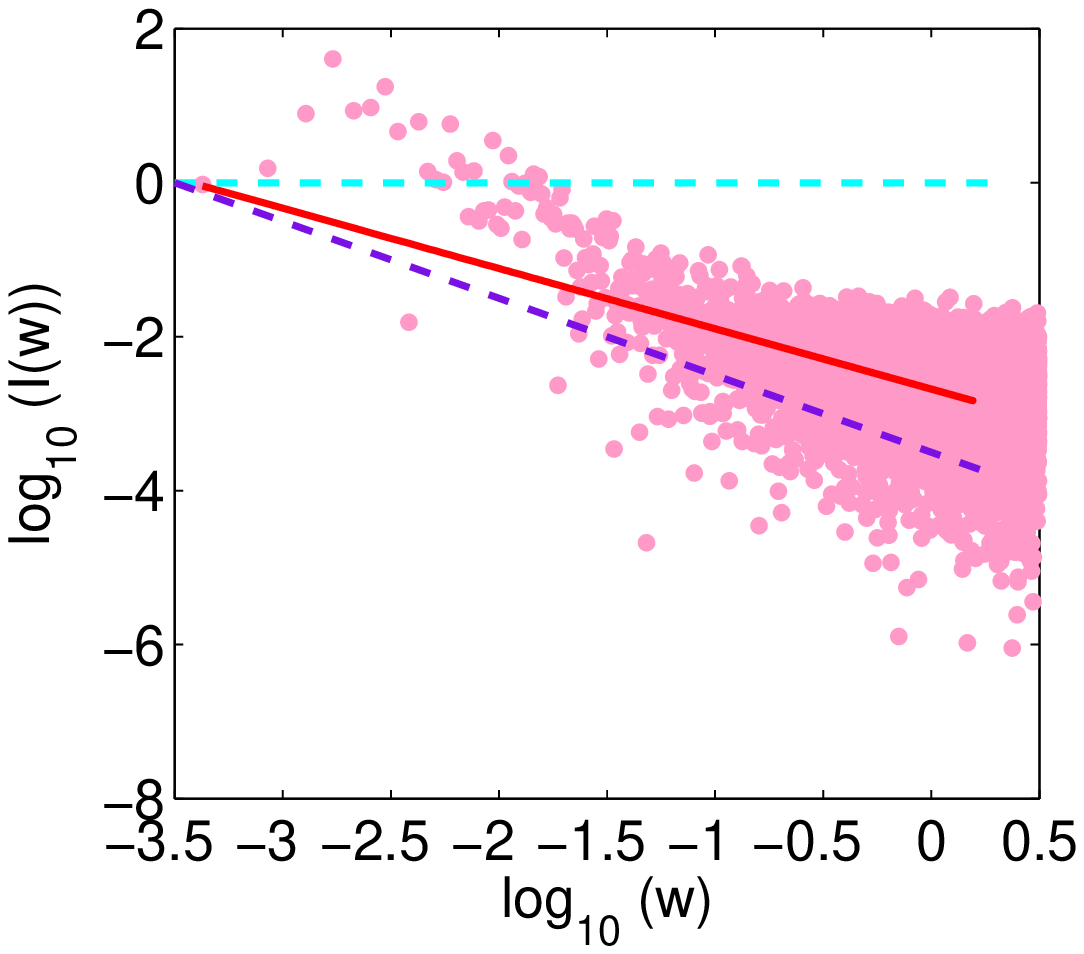}
\end{minipage}
\begin{minipage}[t]{1.8in}
\includegraphics[width=1.8in]{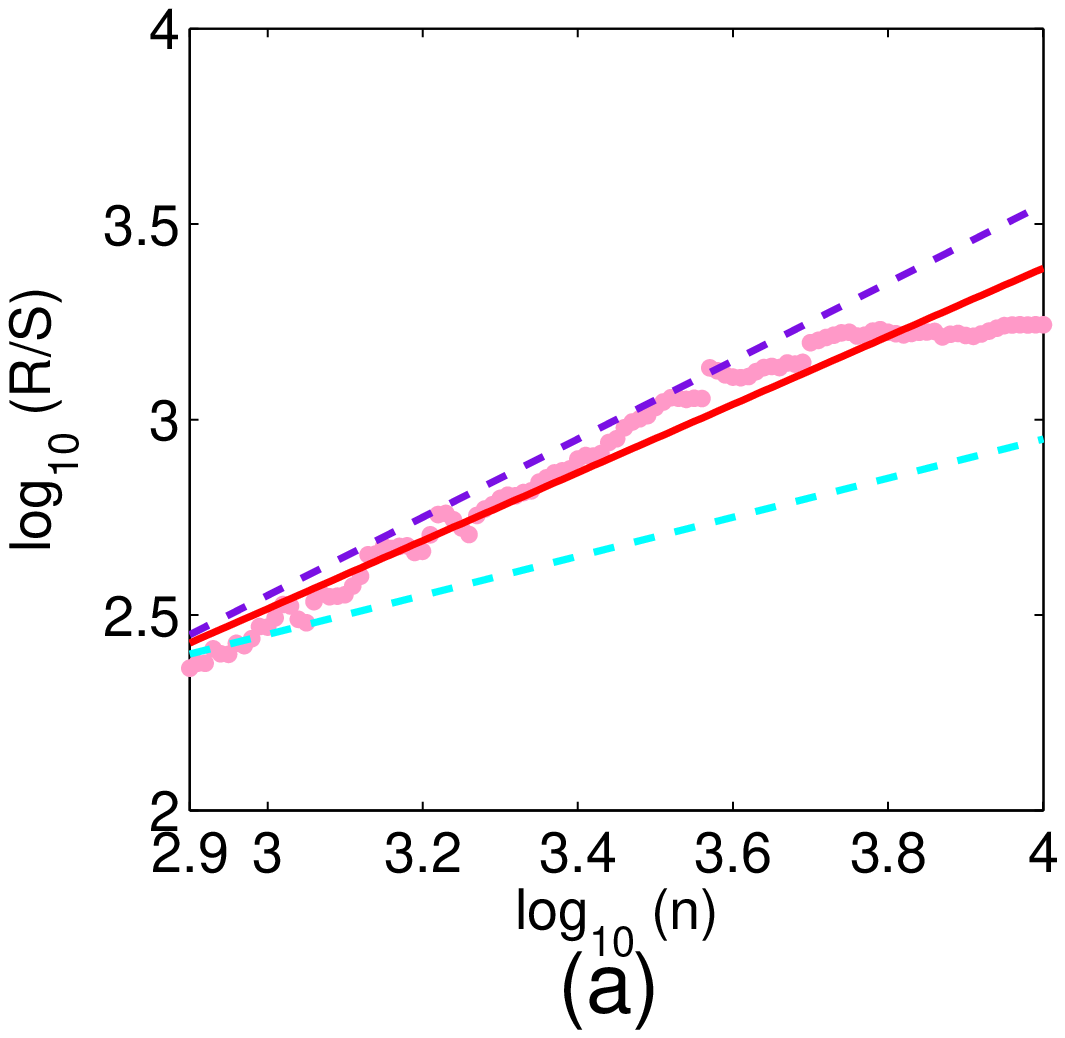}
\end{minipage}
\begin{minipage}[t]{1.8in}
\includegraphics[width=1.8in]{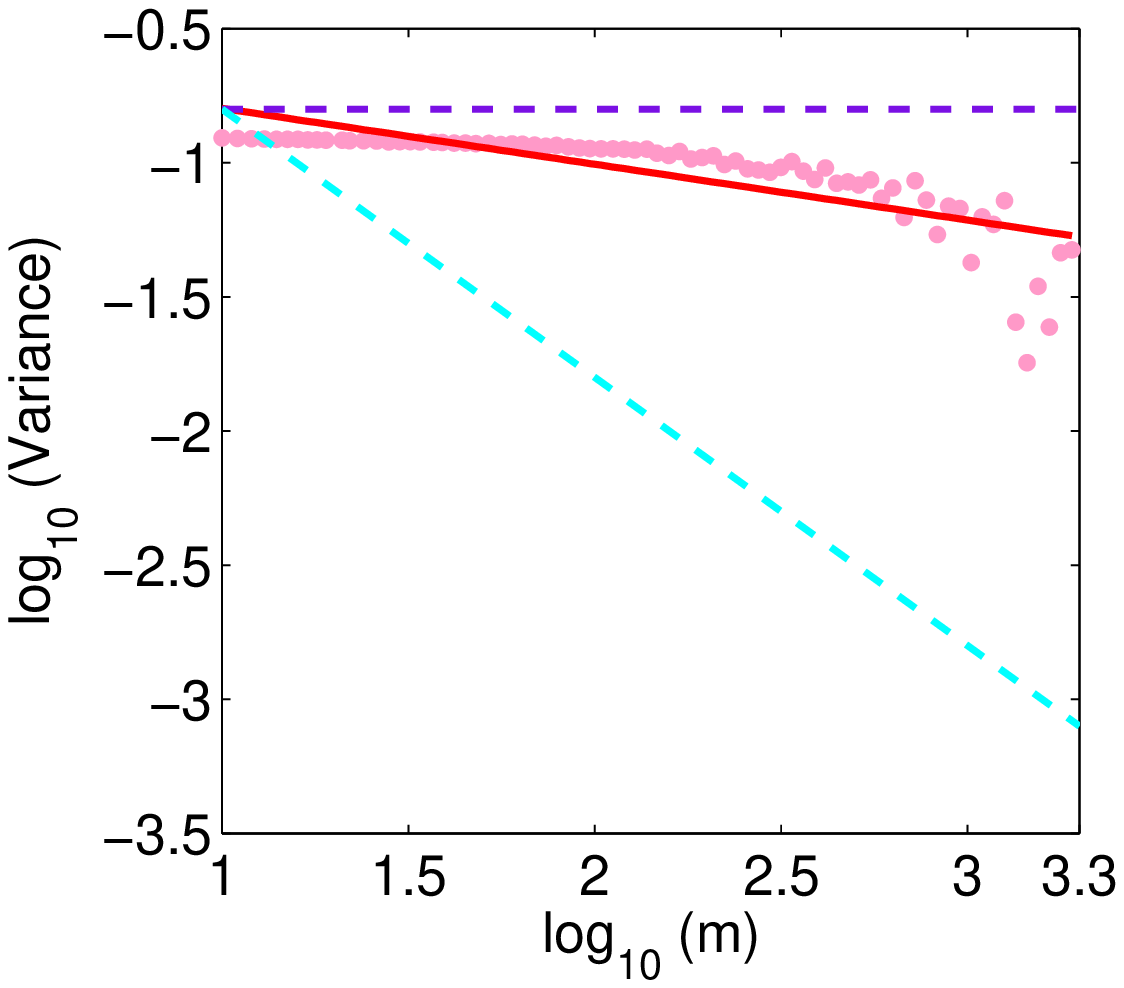}
\end{minipage}\\
\begin{minipage}[t]{1.8in}
\includegraphics[width=1.8in]{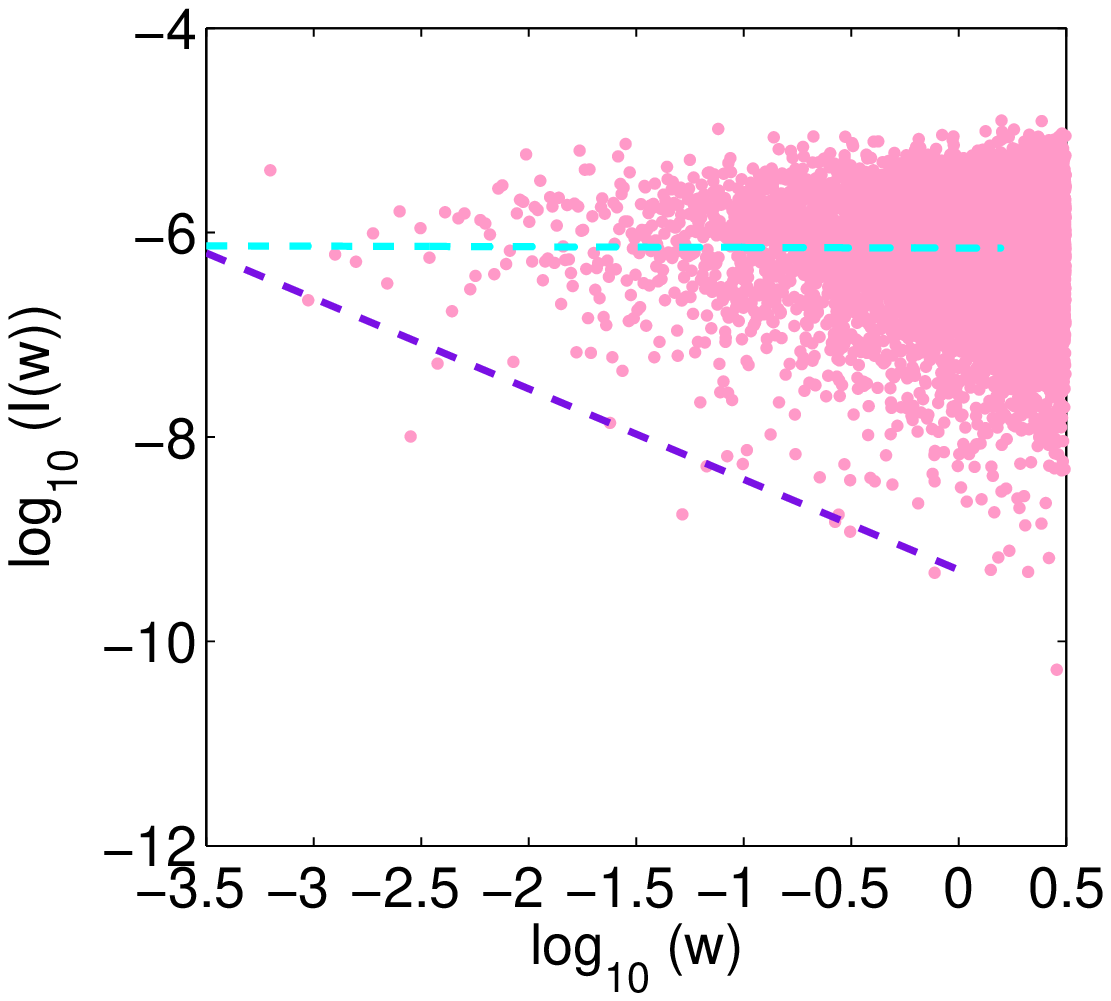}
\end{minipage}
\begin{minipage}[t]{1.8in}
\includegraphics[width=1.8in]{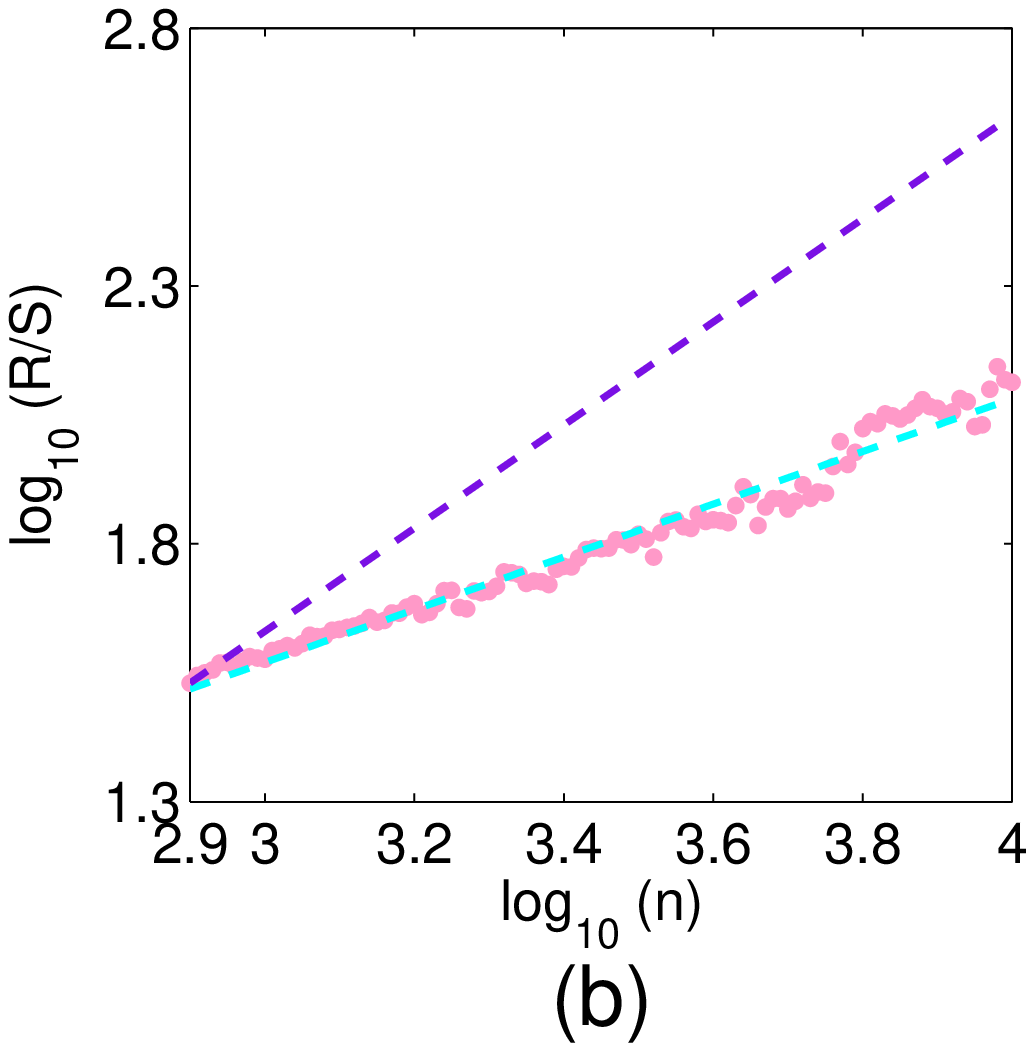}
\end{minipage}
\begin{minipage}[t]{1.8in}
\includegraphics[width=1.8in]{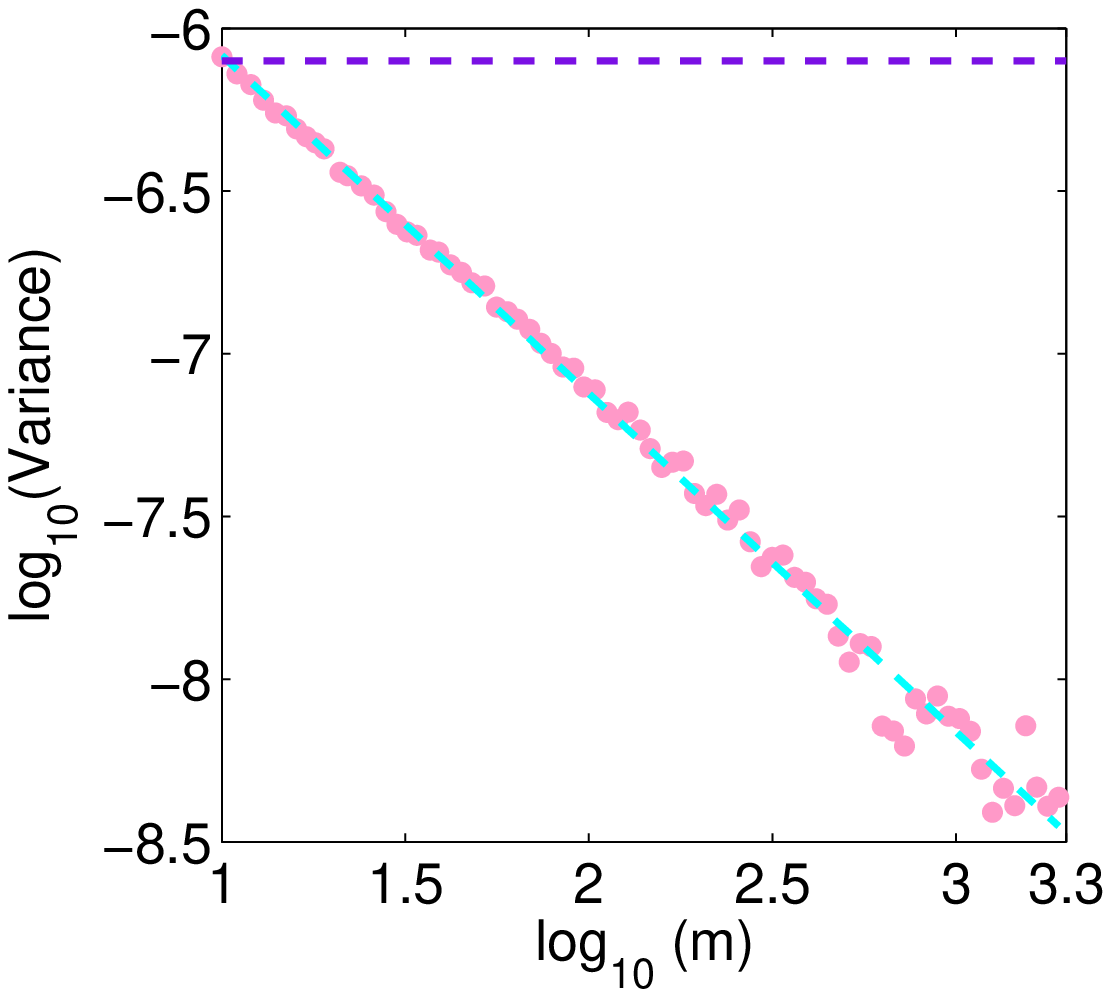}
\end{minipage}
\end{tabular}
\caption{Statistical fractal of the measured and derived wireless cellular coverage boundaries. (a) Statistical fractal estimation of the measured wireless cellular coverage boundary; the red line is the fitted line based on the measured wireless cellular coverage boundary data, where the value of the Hurst parameter is $H \approx 0.9$, the blue broken line corresponds to the value of the Hurst parameter $H = 0.5$, the purple broken line corresponds the value of Hurst parameter $H = 1$. (b) Statistical fractal estimation of the derived wireless cellular coverage boundary; the blue broken line is the fitted line based on the derived wireless cellular coverage boundary data where the value of the Hurst parameter is $H = 0.5$, and the purple broken line corresponds the value of the Hurst parameter $H = 1$. }\label{Fig4}
\end{figure*}

Fig. 4b illustrates the Hurst parameter estimated from the derived wireless cellular coverage boundary data by the periodogram method, R/S method and variance-time analysis method. In the left figure of Fig. 4b, the periodogram method is used to estimate the value of the Hurst parameter by the spectral density of the random process ${X'} = \{ {{\phi}'}_i,\ i = 1,2, \cdots M\}$ in a log-log plot. When the frequency value approaches zero, the spectral density remains nearly constant in the log-log plot. Utilizing the least squares method, the slope, i.e., the decay speed of spectral density in the log-log plot, is estimated to be $\alpha {\rm{ = }}0.001$ . Furthermore, the value of the Hurst parameter is calculated to be $H = \frac{1}{2}(1 + \alpha ) = 0.4995 \approx 0.5$. In the middle figure of Fig. 4b, the R/S method is used to estimate the value of the Hurst parameter based on the R/S statistic of the random process ${X'} = \{ {{\phi}'}_i,\ i = 1,2, \cdots M\}$ in a log-log plot. The R/S statistic still linearly increases with increases in the length of series $n$ in a log-log plot. Compared with the slopes in the middle figure of Figs. 4a and 4b, the slope estimated from the derived wireless cellular coverage boundary is clearly less than the slope estimated from the measured wireless cellular coverage boundary in log-log plots. Utilizing a least squares method, the Hurst parameter, i.e., the rate of increase of the R/S statistic, is estimated to be $H = 0.5011 \approx 0.5$ . In the right figure of Fig. 4b, the variance-time analysis method is used to estimate the value of the Hurst parameter via the variance of aggregated series of the random process ${X'} = \{ {{\phi}'}_i,\ i = 1,2, \cdots M\}$ . The variance of aggregated series presents a linear decaying characteristic with increases in the aggregation level $m$ in a log-log plot. Utilizing a least squares method, the slope of sample variances is estimated to be $\beta {\rm{ = }}0.9690$ in the log-log plot. Furthermore, the value of the Hurst parameter is calculated to be $H = 1 - \frac{\beta }{2} = 0.5155 \approx 0.5$ . Based on the results from the three Hurst parameter estimators, the value of the Hurst parameter estimated from the derived wireless cellular coverage boundary is $H \approx 0.5$.

\begin{table*}[!h]
\setlength{\abovecaptionskip}{0mm}
\centering
\caption{Hurst Parameter of the wireless cellular coverage boundary}\label{tab:2}
{\small
\begin{tabular}{lll}\toprule
\noalign{\smallskip}
\textbf{Hurst Parameter Estimation Method}&\multicolumn{2}{c}{\textbf{Hurst Parameter}}\\
&{Measured wireless cellular }&{Derived wireless cellular }\\
&{coverage boundary}&{coverage boundary}\\ \midrule
\noalign{\smallskip}\noalign{\smallskip}
{Periodogram method} &\multicolumn{1}{c}{0.9013 } & \multicolumn{1}{c}{0.4995}\\
{Rescaled adjusted range statistic method}&\multicolumn{1}{c}{0.9000} &\multicolumn{1}{c}{0.5011}  \\
{Variance-time analysis method}& \multicolumn{1}{c}{0.8898}&\multicolumn{1}{c}{0.5155} \\ \bottomrule
\noalign{\smallskip}
\end{tabular}
}
\end{table*}

Compared with the values of the Hurst parameter in Table \uppercase\expandafter{\romannumeral2}, the values of the Hurst parameter estimated from the measured wireless cellular coverage boundary is located in the interval $\left( {0.5,1.0} \right)$ , i.e., $H \approx 0.9$ . Therefore, the measured wireless cellular coverage boundary has the statistical fractal characteristic in angular scales.

\begin{table*}[!h]
\setlength{\abovecaptionskip}{0pt}
\centering
\caption{Hurst Parameters of other three measured cellular}\label{tab:3}
{\small
\begin{tabular}{llllp{8cm}}\toprule
\noalign{\smallskip}
\textbf{Hurst Parameter Estimation Method} &\multicolumn{3}{c}{\textbf{Hurst Parameter}}\\
&{Zhangjiang Road,}&{Pingjiang Road, }&{Tianshan Road,}\\
&{Shanghai, China.}&{Shanghai, China.}&{Shanghai, China.}\\
&{April 22,2014}&{May 23,2014}&{June 4,2014}\\ \midrule
\noalign{\smallskip}\noalign{\smallskip}
{Periodogram method} &\multicolumn{1}{c}{0.9188} & \multicolumn{1}{c}{0.9120}&\multicolumn{1}{c}{0.9096}\\
{Rescaled adjusted range statistic method} &\multicolumn{1}{c} {0.9313} &\multicolumn{1}{c}{0.9420}&\multicolumn{1}{c}{0.9020}  \\
{Variance-time analysis method}&\multicolumn{1}{c} {0.8865}&\multicolumn{1}{c}{0.8673}&\multicolumn{1}{c}{0.9252}\\
\hline
{Mean value of Hurst}&\multicolumn{1}{c}{0.9122}&\multicolumn{1}{c}{0.9071}&\multicolumn{1}{c}{0.9096}\\ \bottomrule
\noalign{\smallskip}
\end{tabular}
}
\end{table*}

Although the measured wireless cellular data in this study is collected from the Pingjiang Road, Shanghai China, the measurement location was not specially selected for the evaluation of the wireless cellular coverage boundary. Obstacles around the BS are distributed by the city planning, and no specified changes have been forced on the obstacles in the measurement process. Moreover, the measurement route includes the wireless signal propagation fading shaded by office buildings and green belts in urban scenarios. Despite the slight deviation of the Hurst parameter values estimated by the three different typical methods, the final estimated result of the measured wireless cellular coverage boundary can be considered as consistent, i.e., the value of Hurst parameter is $H \approx 0.9$, considering the system error generated by estimators themselves. In addition, we also analyze other measured wireless cellular data from Pingjiang Road on May 23, 2014 and the other two BSs located in the urban and suburban areas of Shanghai, China. The detail analysis results are illustrated in Table \uppercase\expandafter{\romannumeral3}. The analysis results from all of the measured wireless cellular data indicate that the mean values of the Hurst parameter estimated from the measured wireless cellular coverage boundaries approximate 0.9. Therefore, the analysis result in this paper is reasonable, and the real wireless cellular coverage boundary has the statistical fractal characteristic.

\section{Conclusions}
Considering the anisotropy fading of wireless signal propagated in non-free spaces, the statistical characteristics of wireless cellular coverage boundary have been measured and analyzed in this paper. The analyzed results indicate that the measured wireless cellular coverage boundary is extremely irregular and that it is difficult to depict the measured wireless cellular coverage boundary using conventional Euclidean geometry methods. Thus, based on fractal geometry theory, the statistical characteristic of the measured wireless cellular coverage boundary was estimated using three typical Hurst parameter estimators. Our results validate the fact that the real wireless cellular coverage boundary has the statistical fractal characteristic in angular scales. Therefore, real wireless cellular networks can be called wireless fractal wireless cellular networks.

By utilizing fractal geometric theory, random processes which exhibit the fractal characteristic have been put forward to fit wireless fractal cellular networks. Therefore, based on the fractal characteristic validated in this paper, the new system model of cellular networks could be built to analyze and optimize the performance of random cellular networks in the following topics:

\begin{enumerate}
\item Improving the cooperative transmission efficiency. Based on our measured data, the distance between the wireless cellular coverage boundary and the associated BS is more than 180 Km in specified directions. In this case, the new cooperative transmission scheme need to include non-adjacent cooperative BSs located at remote regions and further improve the transmission efficiency considering direction effects among the user and cooperative BSs.
\item Optimizing energy efficiency. The green communication is one of important topics for future cellular networks. Based on results in this paper, a fractal wireless cellular coverage model can be expected to describe the wireless cellular coverage areas. Furthermore, the optimal energy efficiency of cellular networks can be achieved by adjusting the BS transmission power considering wireless fractal cellular coverage areas.
\item Based on the beamforming technologies, new angular power control technologies can be developed to improve the transmission efficiency and energy efficiency of cellular networks. Different from the conventional power control technologies which adjust the BS transmission power based on the channel state information, the new angular power control technologies can adaptively adjust the transmission power at different directions considering the fractal characteristic of wireless cellular coverage boundary.
\end{enumerate}

\section*{Acknowledgments}

The corresponding authors of the article are Prof. Xiaohu Ge and Yang Yang. The authors
would like to acknowledge the support from the International Science
and Technology Cooperation Program of China (Grant No. 2014DFA11640 and 2015DFG12580), the National Natural Science Foundation of
China (NSFC) (Grant No. 61231009 and 61461136003), the NSFC
Major International Joint Research Project (Grant No. 61210002), the China 863 Project (Grant No. 2014AA01A701 and 2014AA01A707), the Science and Technology Commission of Shanghai Municipality (STCSM) under grant 14ZR1439700,
the Fundamental Research Funds for the Central Universities (Grant
No. 2015XJGH011), EU FP7-PEOPLE-IRSES
(Contract/Grant No. 247083, 318992, 612652 and 610524), the EPSRC TOUCAN project (Grant No. EP/L020009/1), and EU H2020 ITN 5G Wireless project (Grant No. 641985).

\begin{IEEEbiography}
{Xiaohu~Ge}
(M'09-SM'11) is currently a full professor with the School of Electronic Information and Communications at Huazhong University of Science and Technology (HUST) China and an adjunct professor with at with the Faculty of Engineering and Information Technology at University of Technology Sydney (UTS), Australia. He received his Ph.D. degree in communication and information engineering from HUST in 2003. He is the director of China National International Joint Research Center of Green Communications and Networks. He has been actively involved in organizing more the ten international conferences since 2005. He served as the general Chair for the 2015 IEEE International Conference on Green Computing and Communications (IEEE GreenCom). He serves as an Associate Editor for the \textit{IEEE ACCESS}, \textit{Wireless Communications and Mobile Computing Journal (Wiley)} and \textit{the International Journal of Communication Systems (Wiley)}, etc.
\end{IEEEbiography}
\vspace{-6 mm}

\begin{IEEEbiography}{Yehong~Qiu}
received her B.E. degrees in Communication Engineering from Huazhong University of Science and Technology (HUST) in 2014. She is currently working towards a Master’s degree in the School of Electronic Information and Communications at HUST. Her research interests mainly include characteristic and performance analysis of wireless fractal cellular networks, energy efficiency of wireless cellular networks.
\end{IEEEbiography}
\vspace{-6 mm}

\begin{IEEEbiography}{Jiaqi~Chen}
 received the B.E. degree in Communication Engineering from Huazhong University of Science and Technology (HUST), Wuhan, China in 2013. She is currently working towards the Ph.D. degree in HUST. Her research interests include energy efficiency, fractal cellular networks, and handover.
\end{IEEEbiography}
\vspace{-6 mm}

\begin{IEEEbiography}{Meidong~Huang}
received the B.E. degree in Communication Engineering from Huazhong University of Science and Technology(HUST), Wuhan, China, in 2015. Now he is working toward the Master's degree in HUST. His research interests mainly in the cellular boundary analysis.
\end{IEEEbiography}
\vspace{-6 mm}

\begin{IEEEbiography}{Hui~Xu}
received the BEng and MEng degrees in Communication Engineering from Shanghai Jiaotong University, Shanghai, P. R. China, in 2004 and 2004, respectively.
Xu Hui is currently a Senior Engineer with Shanghai Institute of Microsystem and Information Technology (SIMIT), Chinese Academy of Sciences, serving as the department head of CAS Key Laboratory of Wireless Sensor Network and Communication, and the department head of Shanghai Research Center for Wireless Communications (WiCO). Prior to that, he has served the Department of Global Telecom Solutions Sector at MOTOROLA (China) co., LTD, as a R\&D Senior Engineer; and the Department of Production Quality at Shanghai Datang Telcom, China, as a R\&D Engineer. His research interests include wireless communication networks, software defined wireless networks, 5G mobile systems, intelligent transport systems, wireless testbed development and practical experiments.
\end{IEEEbiography}
\vspace{-6 mm}

\begin{IEEEbiography}{Jing~Xu}
received his M.S. degrees in electronic engineering from Jilin University, Chang Chun, China, in 2001 and received his Ph.D. degree from the radio engineering department of Southeast University, Nan Jing, China, in 2005. He is a professor with Shanghai Institute of Microsystem and Information Technology (SIMIT) of Chinese Academy of Sciences (CAS), and obtained Shanghai Young Rising Star (2010). His main research interest includes inter-cell interference mitigation,interference modeling, cooperative communications and software defined wireless network.
\end{IEEEbiography}
\vspace{-6 mm}

\begin{IEEEbiography}{Wuxiong~Zhang}
received the B. E. degree in Information Security from Shanghai Jiao Tong University, Shanghai, China, in 2008, and the Ph.D. degree of communication and information system in Shanghai Institute of Microsystem and Information Technology (SIMIT), Chinese Academic of Sciences in 2013. He is currently an assistant professor at SIMIT, and serving as an assistant researcher at Shanghai Research Center for Wireless Communications. His research interests include 4G/5G mobile communication systems and vehicular networks.
\end{IEEEbiography}
\vspace{-6 mm}

\begin{IEEEbiography}{Yang~Yang}
(S'99-M'02-SM'10) received the BEng and MEng degrees in Radio Engineering from Southeast University, Nanjing, P. R. China, in 1996 and 1999, respectively; and the PhD degree in Information Engineering from The Chinese University of Hong Kong in 2002.
Dr. Yang Yang is currently a professor with Shanghai Institute of Microsystem and Information Technology (SIMIT), Chinese Academy of Sciences, serving as the Director of CAS Key Laboratory of Wireless Sensor Network and Communication, and the Director of Shanghai Research Center for Wireless Communications (WiCO). He is also an adjunct professor with the School of Information Science and Technology, ShanghaiTech University. Prior to that, he has served the Department of Electronic and Electrical Engineering at University College London (UCL), United Kingdom, as a Senior Lecturer; the Department of Electronic and Computer Engineering at Brunel University, United Kingdom, as a Lecturer; and the Department of Information Engineering at The Chinese University of Hong Kong as an Assistant Professor. His research interests include wireless ad hoc and sensor networks, software defined wireless networks, 5G mobile systems, intelligent transport systems, wireless testbed development and practical experiments.
Dr. Yang Yang has co-edited a book on heterogeneous cellular networks (2013, Cambridge University Press) and co-authored more than 100 technical papers. He has been serving in the organization teams of about 50 international conferences, e.g. a co-chair of Ad-hoc and Sensor Networking Symposium at IEEE ICC’15, a co-chair of Communication and Information System Security Symposium at IEEE Globecom’15.
Yang's research is partially supported by the National Natural Science Foundation of China (NSFC) under grants 61231009 and 61461136003, the Ministry of Science and Technology (MOST) 863 Hi-Tech Program under grant 2014AA01A707, and the Science and Technology Commission of Shanghai Municipality (STCSM) under grant 14ZR1439700.
\end{IEEEbiography}
\vspace{-6 mm}

\begin{IEEEbiography}{Cheng-Xiang~Wang}
(S'01-M'05-SM'08) received his Ph.D. degree from Aalborg University, Denmark, in 2004. He has been with Heriot-Watt University since 2005 and became a professor in 2011. His research interests include wireless channel modelling and 5G wireless communication networks. He has served or is serving as an Editor or Guest Editor for 11 international journals, including \textit{IEEE Transactions on Vehicular Technology } (2011-), \textit{IEEE Transactions on Communications} (2015-),\textit{ IEEE Transactions on Wireless Communications }  (2007-2009), and\textit{ IEEE Journal on Selected Areas in Communications }. He has published one book and over 240 papers in journals and conferences.
\end{IEEEbiography}
\vspace{-6 mm}

\begin{IEEEbiography}{John Thompson}
 (M'03-SM'12-F'16) is currently a Professor in Signal Processing and Communications at the School of Engineering in the University of Edinburgh. He specializes in antenna array processing, cooperative communications systems and energy efficient wireless communications. He has published in excess of three hundred papers on these topics,including one hundred journal paper publications. He is currently the project coodinator for the EU Marie Curie International Training Network project ADVANTAGE studying smart grid technology and for the EPSRC SERAN research project which investigates concepts for fifth generation wireless systems. He has recently been a distinguished lecturer for the IEEE in the field of Green Communications from 2014-2015. He is an editor for the \textit{Green Communications and Computing Series} that appears regularly in \textit{IEEE Communications Magazine} . In January 2016, he was elevated to Fellow of the IEEE for contributions to to multiple antenna and multi-hop wireless communications.

\end{IEEEbiography}

\end{document}